\documentclass[twocolumn]{aastex63}

\usepackage{graphicx}	
\usepackage{amssymb}	
\usepackage{amsmath}	
\usepackage{mathrsfs}   
\usepackage{grffile} 
\usepackage{xcolor}
\usepackage[caption=false]{subfig}

\newcommand{\xismu}{$\xi_{w}(s,\mu)\,$} 

\def\be{\begin{equation}}
\def\ee{\end{equation}}
\def\ba{\begin{eqnarray}}
\def\ea{\end{eqnarray}}

\def\nn{\nonumber}

\received{xxx}
\revised{xxx}
\accepted{xxx}
\submitjournal{ApJ}

\shorttitle{Distance constraints using BAO peaks only}
\shortauthors{Sridhar et al.}

\begin{document}

\title{The clustering of LRGs in the DECaLS DR8 footprint: distance constraints from baryon 
acoustic oscillations using photometric redshifts}

\correspondingauthor{Srivatsan Sridhar}
\email{srivatsan@kasi.re.kr}

\author[0000-0001-8649-0079]{Srivatsan Sridhar}
\affiliation{Korea Astronomy \& Space Science Institute 776, Daedeokdae-ro, Yuseong-gu, 
	         Daejeon, Republic of Korea (34055)}

\author{Yong-Seon Song}
\affiliation{Korea Astronomy \& Space Science Institute 776, Daedeokdae-ro, Yuseong-gu, 
			 Daejeon, Republic of Korea (34055)}

\author{Ashley J. Ross}
\affiliation{Center for Cosmology and AstroParticle Physics, 
			 The Ohio State University, Columbus, OH 43210}
			 
\author{Rongpu Zhou}
\affiliation{Lawrence Berkeley National Laboratory, 1 Cyclotron Road, Berkeley, CA 94720, USA}

\author{Jeffrey A. Newman}
\affiliation{Department of Physics and Astronomy and PITT PACC, 
		 	 University of Pittsburgh, 3941 O’Hara St., Pittsburgh, PA 15260}

\author{Chia-Hsun Chuang}
\affiliation{Kavli Institute for Particle Astrophysics and Cosmology, Stanford University, 
             452 Lomita Mall, Stanford, CA 94305, USA}

\author{Francisco Prada}
\affiliation{Instituto de Astrof\'{i}sica de Andaluc\'{i}a (CSIC), Glorieta de la Astronom\'{i}a, s/n, E-18008 Granada, Spain}

\author{Robert Blum}
\affiliation{NSF’s Optical–Infrared Astronomy Research Laboratory P.O. Box 26732, Tucson, AZ 85719, USA}

\author{Enrique Gazta\~{n}aga}
\affiliation{Institute of Space Sciences (ICE, CSIC), Campus UAB, Carrer de Can Magrans, s/n, 08193 Barcelona, Spain}
\affiliation{Institut d'Estudis Espacials de Catalunya (IEEC), E-08034 Barcelona, Spain}

\author{Martin Landriau}
\affiliation{Lawrence Berkeley National Laboratory, 1 Cyclotron Road, Berkeley, CA 94720, USA}



\begin{abstract}

A photometric redshift sample of Luminous Red Galaxies (hereafter LRGs) obtained from The 
DECam Legacy Survey (DECaLS) is analysed to probe cosmic distances by exploiting the 
wedge approach of the
two-point correlation function. Although the cosmological information is highly contaminated 
by the uncertainties existing in the photometric redshifts from the galaxy map, an angular 
diameter 
distance can be probed at the perpendicular configuration in which the measured correlation 
function is minimally contaminated. An ensemble of wedged correlation functions selected up 
to a given threshold based on having the least 
contamination was studied in the previous work \citep{Srivatsan_2019} using simulations, 
and the extracted cosmological information 
was unbiased within this threshold. We apply the same methodology 
for analysing the LRG sample from DECaLS which will provide the optical imaging for targeting
two-thirds of the DESI footprint and measure the angular diameter distances at $z=0.69$ and 
$z=0.87$ to be $D_{A}(0.697)=(1499 \pm 77\,\mathrm{Mpc})(r_{d}/r_{d,fid})$ and 
$D_{A}(0.874)=(1680 \pm 109\,\mathrm{Mpc})(r_{d}/r_{d,fid})$ with a fractional error of 
5.14\% and
6.48\% respectively. We obtain a value of $H_{0}=67.59\pm5.52$ km/s/Mpc which supports
the $H_0$ measured by all other BAO results and is consistent with $\Lambda$CDM model.

\end{abstract}



\section{Introduction} \label{sec:intro}

Measuring the expansion history of the Universe is of paramount importance in the field of modern 
cosmology. It can be revealed by diverse cosmic distance measures in tomographic redshift space, 
such as cosmic parallax~\citep{Parallax}, standard candles~\citep{Standard_candle} or 
standard rulers~\citep{Eisenstein_1998,Eisenstein_2005}. To date the best constraints come from 
the distance-redshift relation and imply that the expansion rate has changed from a decelerating phase
to an accelerated one \citep{Riess_1998,Perlmutter_1999}. As most ongoing observations support 
the $\Lambda$CDM model with the presence of the cosmological constant, but to confirm it with high 
precision or to possibly find any deviation from it still remains an interesting observational 
mission. One of the most robust methods for measuring distance-redshift relation is to use the 
baryon acoustic oscillation feature that is observed as a bump in the two-point correlation 
function or as wiggles in the power spectrum. 
The tension between gravitational infall and radiative pressure caused by the 
baryon-photon fluid in the early Universe gave rise to an acoustic peak structure which was imprinted
on the last-scattering surface (hereafter BAO) \citep{Peebles_Yu}. The BAO feature has been measured
through the correlation function \citep{Eisenstein_2005}, and the most successful 
measurements in the clustering of large-scale structure at low redshifts have been obtained
using data from SDSS \citep{Eisenstein_2005,Estrada_2009,Padmanabhan_2012,Hong_2012,Veropalumbo_2014,Veropalumbo_2016,Alam_2017}. 
The Dark Energy Spectroscopic Instrument (DESI) is an upcoming 
survey \citep{DESI} which will be launched to probe the earlier expansion history with greater 
precision using spectroscopic redshifts. However, the photometric footprint for DESI has already 
been completed by the Legacy Imaging Surveys \citep{Decals}.  
Photometric surveys provide more observed galaxies compared to a spectroscopic survey even 
at deeper redshifts \citep{Euclid_Srivatsan_2019}, but the uncertainty on the redshift 
obtained from photometric surveys is larger compared to the uncertainty on the redshift 
obtained from spectroscopic surveys. 
Although these photometric redshifts are measured with a much poorer resolution and an 
unpredictable damping of clustering at small scales and a smearing of the BAO peak is caused 
by the photo-z uncertainty~\citep{Estrada_2009}, possible BAO signatures that have not 
been washed-out by the redshift uncertainty might still be present. 

We investigate the optimised methodology to extract the 
cosmic distance information from the 
photometric datasets and provide a precursor of cosmic distance information which will be 
revealed by the follow up spectroscopy experiment much later on. We apply the wedge approach 
\citep{Kazin_2013,Sanchez_2013,Sanchez_2014,Cris_2016,Ross_2017,Sanchez_2017,Srivatsan_2019} 
to probe the uncontaminated 
BAO feature by binning the angular direction from the perpendicular to radial directions, and 
recover the residual BAO peak that has survived and get constraints on 
the angular diameter distance $D_{A}$ and $H^{-1}$. 
It has also been shown recently by \cite{Ross_2017} that the statistics obtained using 
wedge correlation function are about 6\% more 
accurate compared to the angular correlation function. Thus, using \xismu not only adds more 
information compared to $w(\theta)$, but also overcomes the above disadvantages.

Recently, some improved methodologies have measured the Hubble constant in great precision, 
which reveal a tension among measurements. This tension draws attention to the community as a 
possible presence of new physics or unknown systematic uncertainties that need to be fixed. 
The Hubble constant is indirectly measured by the highest resolution cosmic microwave background 
maps provided by the Planck satellite experiment \citep{Planck_main} and they find it to be 
$H_0=67.4\pm0.5 {\rm\,km/s/Mpc}$ \citep{Planck_2018}. The Hubble constant can also be directly 
probed by classical distance ladder using type Ia Supernovae samples 
\citep{Scolnic_Srivatsan_supernovae}. The latest value from 
\cite{Riess_2019} give us a constraint of $H_0=74.03 \pm 1.42 {\rm\,km/s/Mpc}$ with a 
few percent marginal 
error. Both efforts leaves a huge discrepancy in the $H_0$ measurement, with the values being 
4$\sigma$ apart, which needs to be resolved.

While the current analyses of most cosmological observations at low redshift support the 
$H_0$ measured by Planck, next generation survey programs such as DESI will be launched in the 
near future. DESI will probe the earlier expansion history with 
greater precision using spectroscopic redshifts. However, by using the DECaLS data, which provides 
us constraints on the angular diameter distance and by using the information of the sound 
horizon from Planck, we get constraints on $H_{0}$. 
Our analyses uses a fiducial cosmological model with the following parameters: 
$\Omega_{m} = 0.31$, $\Omega_{b} = 0.049$, 
$h \equiv H_{0}/(100\,\mathrm{km s^{-1} Mpc^{-1}}) = 0.676$, $n_{s} = 0.96$ and 
$\sigma_{8} = 0.8$. The paper is organised as follows. In Section \ref{sec:data}, we describe the 
DECaLS DR8 data data including the magnitude cuts we employ for our sample. Section \ref{sec:methodology}
describes the clustering measurements and the fitting procedure used. We present our cosmic distance
constraints obtained in Section \ref{sec:measured_cosmic_distance} and discuss our overall results
and conclusions in Section \ref{sec:conclusion}.

\section{The Data} \label{sec:data}

In this section, we describe the DESI Legacy Imaging Survey DR8 data used in this paper 
along with the Dark Sky simulation data used for testing and validating our results.

\subsection{DECaLS DR8 data}\label{sec:decals_data}

The DESI Legacy Imaging Surveys will provide the target catalogue for the 
upcoming DESI survey. One among the 3 imaging projects conducted for the Legacy Survey 
is DECaLS \citep{Decals} which covers the South Galactic Cap region at DEC $\leq$ 34$^{\circ}$. 
The data makes use of  
three optical bands (\textit{g,r,} and \textit{z}) to a depth of at least $g=24.0$, $r=23.4$ 
and $z=22.5$, which is 1-2 magnitude deeper than SDSS. 
We use the DECaLS data from the Legacy Surveys eighth data release (DR8), which
is the first release to include images and catalogues from all three of the Legacy Surveys in a 
single release. The Legacy Surveys also processed some of the imaging data from the Dark Energy Survey (DES, \citealt{DES}), and we include the DES imaging with DEC $\geq$ -30$^{\circ}$ in our analysis.
In addition to the optical imaging, 4 years of Wide-Field Infrared Survey
Explorer (WISE) \citep{WISE,Wise_meisner} data in the W1 and W2 bands are also included, 
which provide additional colour information.

For the parent LRG sample in this study, we use a non-stellar cut of 
$(z-\mathrm{W}1)-0.8*(r-z)>-0.6$, a faint limit of $z<20.41$, a colour cut 
of $0.75 < (r-z) < 2.45$ and a sliding magnitude-color cut
of $(z-17.18)/2 < (r-z)$. These selection cuts are motivated by the current DESI LRG 
target selection cuts \citep{DESI}. 

We also apply masks to get the final footprint for our parent sample using the ``MASKBITS'' column 
in the DR8 catalog \footnote{\url{http://legacysurvey.org/dr8/bitmasks/\#maskbits}}. Objects 
(and randoms) with following bits are removed: 1 (Tycho-2 and GAIA bright stars), 8 (WISE W1 
bright stars), 9 (WISE W2 bright stars), 11 (fainter GAIA stars), 12 (large galaxies) and 13 
(globular clusters).
Imaging datasets often suffer from systematic effects, 
and one such major contribution towards the systematic contamination comes from correlation with 
stellar density \citep{Decals_2019}. 
A more detailed test on this effect on the large-scale structure correlation 
is explained in Section \ref{sec:appendix1}.

\begin{figure*}
    \centering
	\includegraphics[width=0.47\textwidth,height=3in]{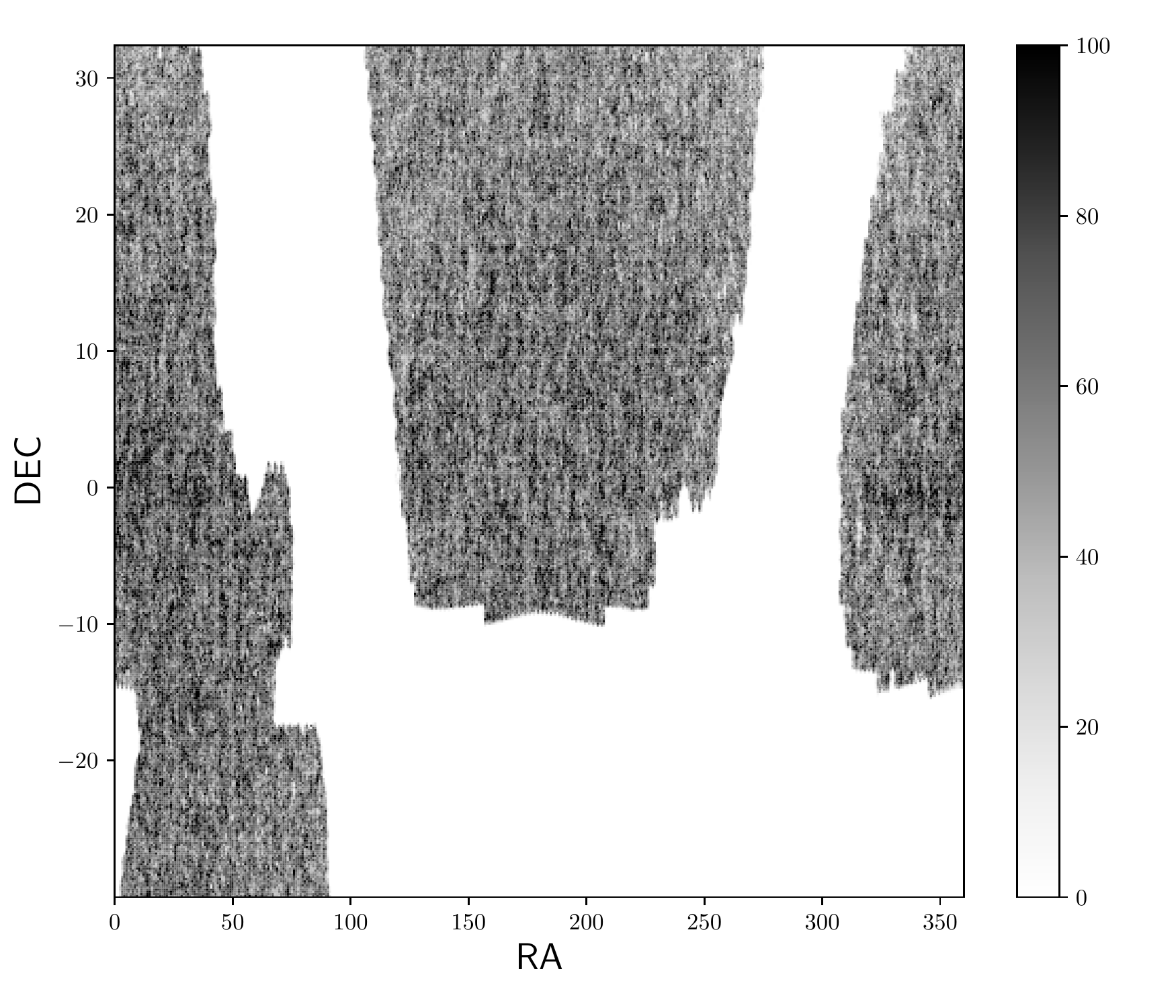}  
	\includegraphics[width=0.47\textwidth,height=3in]{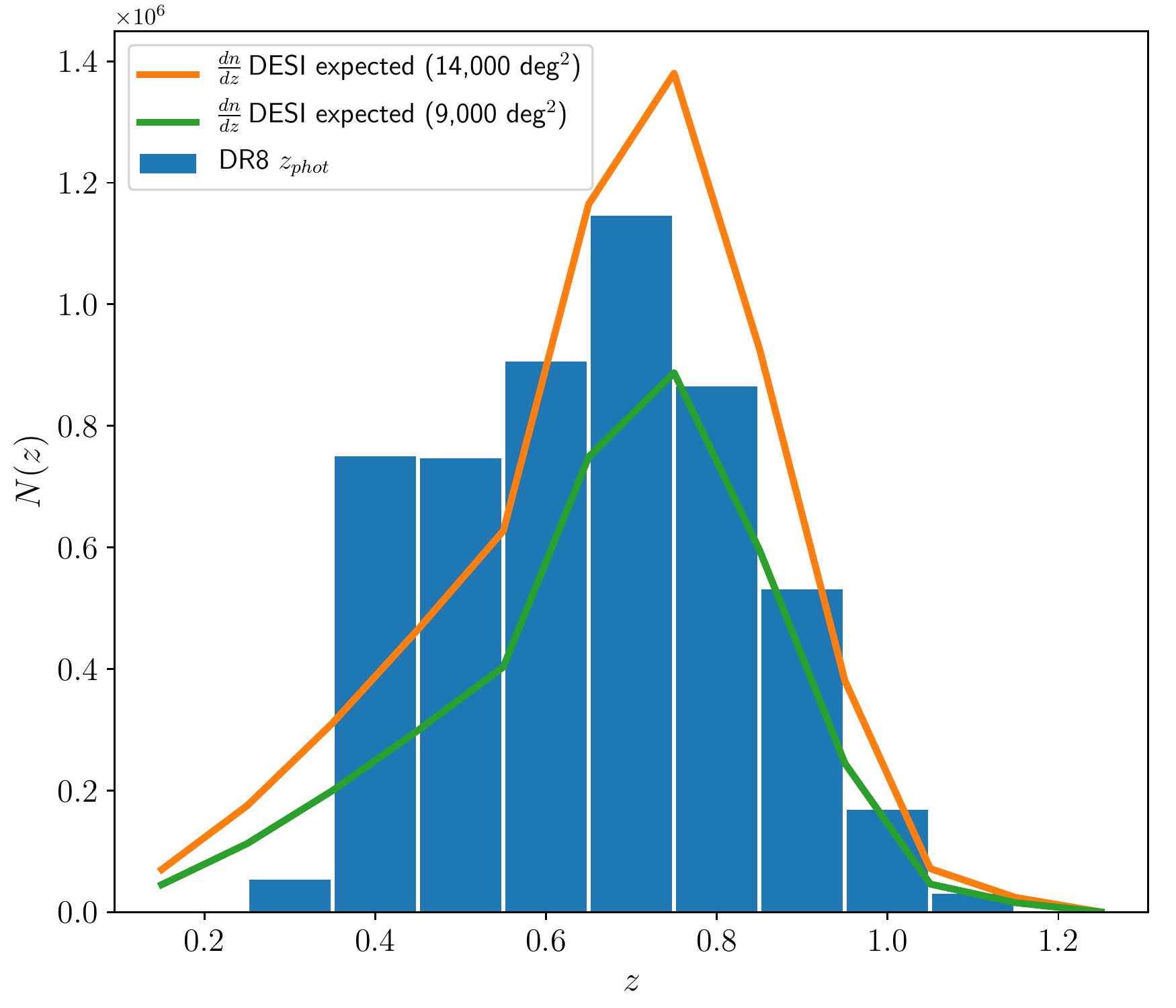} 
	\caption{\textit{Left panel:} The angular distribution of the LRGs from the DECaLS DR8 
	         photometric sample after applying the masking scheme and selection cuts. The density 
	         variations are shown using a normalised ``grayscale'' colormap, 
	         with darker regions denoting the high dense regions. 
	         \textit{Right panel:} The photometric redshift distribution of the DECaLS DR8 
	         parent sample (blue filled histogram) obtained using the random forest method. 
	         The green and orange solid lines
	         are the forecasted $dN/dz$ for LRG galaxies achievable by DESI  
	         \citep{DESI} for a sky coverage of 9000 and 14000 deg$^{2}$ respectively.}
	\label{fig:RA_vs_DEC}   
\end{figure*}

After applying the magnitude cuts and masking scheme, 
we use random forest-based \citep{Breiman2001} photo-z's from \cite{Rongpu} 
to obtain the final photometric redshifts. 
The dispersion on the redshift is usually approximated by, 
\begin{equation}\label{eqn:sigma_z}
\sigma_{z} \equiv \sigma_{0}\times(1+z_{true}) ,
\end{equation}
where $\sigma_{0}$ denotes the dispersion at redshift $z=0$ and $z_{true}$ is the true redshift 
or the spectroscopic redshift. The mean redshift 
uncertainty for the DECaLS DR8 sample within the range $0.3<z_{phot}<1.2$ is $\sigma_{0}=0.0264$. 
The angular distribution of the parent LRG sample after applying the masking and selection cuts is 
plotted in the left panel of Figure \ref{fig:RA_vs_DEC}, with higher density regions denoted by a 
darker shade. The redshift distribution of the parent sample within the range $0.3<z_{phot}<1.2$ 
is plotted as blue filled histogram in the right panel of Figure \ref{fig:RA_vs_DEC}. For comparison, 
we also overplot the forecasted $dN/dz$ LRG redshift distribution from DESI for a sky coverage of
9000 deg$^{2}$ (given by the green solid line) and 14000 deg$^{2}$ (given by the orange solid line). 
The DECaLS DR8 sample has a sky coverage of $\approx$ 9500 deg$^{2}$, which more than two-thirds of 
the 14,000 deg$^{2}$ DESI footprint.

\subsection{Dark Sky simulation data}

In order to test and validate the results we obtain, we also need to analyse the 
clustering from a realistic mock catalogue based on numerical simulations that calculate the 
non-linear evolution of structure and predict the dependence of survey observables on 
cosmological parameters. We use the publicly available Dark Sky simulation set as described in 
\cite{Darksky_2014} for this purpose. The simulation has been generated using particle numbers 
varying from 2048$^{3}$ to 10240$^{3}$ and in a comoving cosmological volume varying from 
$100$h$^{-1}$Mpc to $8$h$^{-1}$Gpc box on a side. The objects have been placed in the simulation 
using a simple (time-evolving) Halo Occupation Distribution (HOD) assuming spherical, 
Navarro-Frenk-White \citep{NFW}, halos for the satellite. To identify dark matter halos and 
substructures, the ROCKSTAR halo finder \citep{rockstar}
has been used. The halo-finding approach is based on an adaptive hierarchical refinement of
friends-of-friends groups in both position and velocity. 
From the set of simulations (with varying particle numbers), we make use of the {\tt ds14\_a}
simulation which has 10240$^{3}$ particles and a particle mass of $3.9\times10^{10}h^{-1}M_{\odot}$. 
These specific mocks contain only RA, DEC and true redshift information for objects pre-identified 
to be LRGs and do not contain color or luminosity information. 

To validate our results on photometric redshifts, we generate a set of photo-z's 
using the true redshift information available. 
In reality, the statistical nature of the photo-z error is more complicated to be 
specified with any known distribution function, but it is assumed that the error propagation of
photo-z uncertainty into cosmological information is mainly caused by the dispersion length
\citep{pablo_arnalte}. 
Thus a simple Gaussian function of statistical distribution is chosen for generating the 
photo-z uncertainty distribution, and we apply the photo-z error dispersion 
$\sigma_z$ as defined in Eq. \ref{eqn:sigma_z} (a more detailed description is given in Section 
\ref{sec:decals_and_darksky_measurements}). In reality, the 
precision is dependent on many factors such as magnitude and spectral type, but here only 
the redshift factor is counted in Eq.~\ref{eqn:sigma_z} in which the coherent 
statistical property determined only by $z$ is applied for all types of galaxies in the 
simulation. 

Table \ref{tab:table1} summarises the number density information from 
both the DECaLS data and the Dark Sky data for 
the entire redshift range and for the two redshift cuts used in this paper. 
It can be seen that 
the number density in all the three redshift ranges for the mock is smaller than the DECaLS sample. 
Thus, the comparison in the clustering between the two samples should only be looked into as a 
consistency check rather than a precise validation of our results.

\begin{table}
\begin{center}
\begin{tabular}{c c c c c} 
\hline\hline
 & Redshift range & $N_{gals}$ & $V$ (Gpc$^{3}$) & $\sigma_{0}$ \\ [0.5ex] 
\hline
 & $0.3<z_{phot}<1.2$ & 5193078 & 7.64 & 0.0263\\ 
DECaLS & $0.6<z_{phot}<0.8$ & 2083394 & 1.74 & 0.0262\\
 & $0.8<z_{phot}<1.0$ & 1074916 & 2.25 & 0.0352\\
\hline
 & $0.3<z_{phot}<1.2$ & 2781896 & 6.51 & 0.0284\\ 
Dark Sky & $0.6<z_{phot}<0.8$ & 1271670 & 1.65 & 0.0282 \\
 & $0.8<z_{phot}<1.0$ & 560749 & 2.16 & 0.0341 \\ [1ex]
\hline
EZmock & $0.6<z_{phot}<0.8$ & 2058906 & 1.65 & 0.0263 \\
 & $0.8<z_{phot}<1.0$ & 963673 & 2.16 & 0.0342\\ [1ex]
\hline
\hline
\end{tabular}
\caption{Number of LRGs, volume for the sample within the range $0.3<z_{phot}<1.2$ (also for the 
         two redshift cut samples used in this paper) and the mean redshift uncertainty within the 
         redshift range ($\sigma_{0}$) for the DECaLS, Dark Sky mock and the EZmock sample 
         (average values from the realisations). 
         The $N_{gals}$ quoted is the total number of LRGs used in the large-scale clustering
         analysis.}
\label{tab:table1}
\end{center}
\end{table}

\subsection{EZmock simulation data for covariance matrix}\label{sec:ezmock_data}
To compute the error on $\xi_w(s,\mu_i)$, we make use of 100 EZmock \citep{EZmock}
simulations all of which have the DESI expected sky coverage. To match the DECaLS DR8 footprint, 
we cut the EZmock samples within  $-30^{\circ}< \mathrm{DEC} < $ 34$^{\circ}$. The EZmock sample
after the DEC cut has an area of $\approx$ 9300 deg$^{2}$, which is similar to the are of the 
DECaLS DR8 parent sample that we use in this paper. 
The mocks contain RA, DEC, $z_{cosmo}$ and $dz_{rsd}$ information. 
Thus, we need to generate photometric redshifts for the mocks so that they can be used to 
obtain the covariance matrix, and we do so using the Gaussian approximation 
following Eq. \ref{eqn:sigma_z}.

To generate the photo-z's for our sample, 
we obtain $\sigma_{z}$'s randomly from the parent DECaLS sample, 
but by restricting to galaxies of similar redshifts that are within a redshift range of $\pm$0.1
$z_{phot}$ which ensures that the dependence of errors on redshift is included. 
For example, for $N$ number of EZmock galaxies that are within $0.6<z_{cosmo}<0.7$, 
$N$ $\sigma_{z}$'s from the DECaLS data within $0.6<z_{phot}<0.7$ are randomly selected. This process
is repeated over the entire redshift range. 
Once we generate the photo-z's, they are diluted according to the DECaLS $N(z)$ to make
sure that they are consistent. The number of galaxies, volume within the redshift range and 
the $\sigma_{0}$ for the two redshift cuts is mentioned in Table \ref{tab:table1}. 
Our detailed analysis comparing the \xismu between the EZmock sample and the DECaLS sample are  
explained in Section \ref{sec:appendix2} and shown in Figure \ref{fig:ezmock_diff_redshift_xismu}.

\section{Methodology}\label{sec:methodology}

In this paper, we follow the same 
methodology and formulation that was applied in \cite{Srivatsan_2019} to simulated 
photometric galaxy catalogues to get cosmological distance constraints. We explain in detail the 
clustering measurements obtained from the wedge correlation function and the comparison between 
the DECaLS and Dark Sky mock data.

\subsection{Clustering measurements and fitting procedure}\label{sec:fitting_procedure}

The excess probability of finding two objects relative to a Poisson distribution at volumes 
$dV_{1}$ and $dV_{2}$ separated by a vector distance \textbf{r} is given by the two-point 
correlation function $\xi(r)$~\citep{totsuji_1969,davis_peebles_1983}. The galaxy 
distribution seen in redshift space exhibits an anisotropic feature distorting 
$\xi(r)$ into $\xi(\sigma,\pi)$ along the line-of-sight (LOS) where $\sigma$ and $\pi$ denote the 
transverse and radial components of the 
separation vector \textbf{r}. Acoustic fluctuations of the 
baryon--radiation plasma of the primordial Universe leaves the signature on the density 
perturbation of baryons. 
This standard ruler length scale, set by the acoustic wave, propagates until 
it is frozen at decoupling epoch to remain in the large scale structure of the Universe. The 
threshold length scale of the acoustic wave is called the sound horizon, which is given by,
\begin{equation} 
r_{d} = \int_{z_{drag}}^{\infty} \frac{c_{s}(z)}{H(z)}dz \end{equation}
where $c_{s}$ is sound speed of the plasma.
This scale is imprinted on the correlation function as a peak and is imprinted on the matter power 
spectrum as a series of waves. Assuming standard matter and radiation content in 
the Universe, the \cite{Planck_2018} measurements of the matter and baryon density determine the 
sound horizon to 0.2\%. By measuring the BAO feature using an anisotropic analysis, one can 
separately measure $D_{A}(z)$ and $H^{-1}(z)$. But adjustments to the cosmological parameters
or changes to the pre-recombination energy density can alter the value of 
$r_{d}$ \citep{Alam_2017}. So, the BAO 
measurements constrain the combinations $D_{A}(z)/r_{d}$, $H^{-1}(z)r_{d}$.  
The sound horizon for this fiducial model is $r_{d,fid}=147.21$ Mpc 
as obtained from \cite{Planck_2018}. The
scalings of $r_{d}$ with cosmological parameters can be found in detail in \cite{Aubourg_2015}.
The distance constraints quoted in this paper are in units of Mpc and with a scaling factor, 
e.g., $D_{A}(z)\times(r_{d,fid}/r_{d})$, so that the numbers provided are independent of the 
fiducial cosmological parameters used.

The Landy \& Szalay estimator (hereafter LS) in $(s,\mu)$ coordinates is best suited to calculate
the two-point correlation function \citep{Farrow,Srivatsan} and extract BAO information from 
photometric redshift galaxy maps \citep{Srivatsan_2019}. 
The radius to shell $s$ and the observed cosine of the angle the galaxy 
pair makes with respect to the LOS $\mu$ are given by
$s^2=\sigma^2+\pi^2$ and $\mu=\pi/s$ respectively, where $\sigma$ and $\pi$ denote the 
transverse and radial separation between the galaxy pairs. 

It is common practice to separate the random sample distributions into the angular and redshift 
components separately. We make use of the random catalogue provided in 
the DR8 data release \footnote{\url{http://legacysurvey.org/dr8/files/}}
by the Legacy Survey, which gives us the angular component. These 
randoms have been downsampled to the surface density of 10000 $/$deg $^{2}$ and requiring +2 
exposures in $g$, $r$ and $z$ bands. The number of objects are usually twice 
or more than the data catalogue to avoid shot noise effects. In our case the random 
catalogue has 5 times more objects than the data catalogue. For the redshift component, 
we extract redshifts randomly from the data catalogue within the chosen redshift range
\citep[see][for more info]{Ross_2012,Ross_2017,Veropalumbo_2016, Srivatsan_2019}. 
The same number of exposure requirements, footprint cuts,
and bright star masks are applied on the randoms as used in constructing the LRG sample.
We use the publicly available {\tt KSTAT} (KD-tree Statistics Package) code \citep{KSTAT} to 
calculate all our correlation functions.

We pay attention to the usefulness of exploiting the wedge correlation function to 
separate the radial contamination from the BAO signal imprinted on perpendicular 
configuration pairs. The wedge correlation function $\xi_w$ is given by,
\begin{equation}\label{eqn:wed_xismu}
\xi_w(s,\mu_i) = \int^{\mu_i^{\rm max}}_{\mu_i^{\rm min}} d\mu'W(\mu':\mu_{\rm cut}=\mu_i^{\rm max})\xi(s,\mu') ,  
\end{equation}
where $\mu_i$ is the mean $\mu$ in each bin (we will refer to the mean value of the $\mu$ bin using
$\bar \mu$ hereafter), 
and $\mu_i^{\rm min}$ and $\mu_i^{\rm max}$ 
are the minimum and maximum values of $\mu$, and $W$ is a window function within the chosen minimum 
and maximum limits of $\mu$. 
Using too many $\mu$ bins will complicate the covariance matrix and by using very few $\mu$
bins we will not be able to separate the error propagation along the LOS clearly \citep{Cris_2016}. 
Thus, we choose 6 bins in the $\mu$ direction 
with $\Delta \mu=0.17$ between $\mu=0$ and 1 with $\mu \rightarrow$ 0 corresponding to 
the transverse plane and $\mu \rightarrow$ 1 corresponding to the LOS plane.

In the case of photometric redshift samples, the noise on the pairs increases along the radial 
configuration and thus causes a smearing of the BAO peak \citep{Estrada_2009}. 
This smearing not only increases with increasing photometric uncertainty but also increases along 
the LOS for a given $\sigma_{0}$. These noisy pairs can be removed	
using a cutoff $\bar \mu$. It has been shown in \cite{Srivatsan_2019} that using a
cutoff $\bar \mu = 0.42$ for photometric redshift samples can remove most of the 
contaminated pairs and thus we use a cutoff $\bar \mu = 0.42$ in this paper.
The empirical model that we use to fit the correlation function and obtain the BAO peak 
location is similar to the one proposed by \citet{Sanchez_empirical,Sanchez_empirical2} 
and is given by,
\begin{equation}\label{eqn:empirical_fit}
\xi_{\rm mod}(s) = B + \left( \frac{s}{s_{0}} \right) ^{- \gamma} + \frac{N}{\sqrt{2\pi \sigma^{2}}} \mathrm{exp} \left( -\frac{(s-s_{m})^{2}}{2\sigma^{2}} \right) ,
\end{equation}
where $B$ takes into account a possible negative correlation at very large scales, 
$s_{0}$ is the correlation length (the scale at which the correlation function $\simeq$ 1) 
and $\gamma$ denotes the slope. The remaining three parameters, 
$N$, $\sigma$ and $s_{m}$ are the parameters of the Gaussian function that model the BAO 
feature and, in particular, $s_{m}$ represents the estimate of the BAO peak position.
This empirical model can be used to accurately extract the BAO peak 
position \citep{Sanchez_empirical,Veropalumbo_2016,Srivatsan_2019} when the correlation function 
is provided.

The likelihood on $s_{m}$ from previous BAO studies is either obtained by using a 1d grid on $s_{m}$
where the $\chi^{2}$ is minimized at each grid point or from the marginalized posterior from a 
Monte Carlo Markov Chain (MCMC) analysis. In this study, the fitting is performed by applying the 
MCMC technique (we make use of the {\tt emcee} Python package \citep{emcee}) , 
using the full covariance matrix obtained using Eq.\ref{eqn:cov_matrix}. 
The fitting parameter space is given by, 
\begin{equation}\label{eqn:param_space}
    x_p=(B,s_0,\gamma,N,s_m,\sigma)
\end{equation} 
and we place flat, wide priors on all the 6 parameters. For the first three parameters, the range 
of the priors are $0.0<B<1.0$, $0.0<s_{0}<3.0$ and $0.0<\gamma<3.0$ and for the remaining 
three parameters
of the Gaussian function, the range of the priors are $0.0<N<1.0$, $85.0<s_{m}<130.0$ and 
$0.0<\sigma<35.0$. Several variations of the range of these priors were tested, especially the range
of the prior on $s_{m}$ and $\sigma$. A smaller range for the $s_{m}$ affects posterior distribution 
and we miss most of the information at high $s(h^{-1}\mathrm{Mpc})$. A similar effect is seen when 
we use a smaller range for the $\sigma$ prior, which eventually amplifies the BAO peak. 
We have also tested several ranges within which to perform the fit and we
fit the correlation function within the range $30.0<s(h^{-1}\mathrm{Mpc})<130.0$ after experimenting
with other ranges. We find that there is a maximum shift of 1\% in the BAO peak when we   
vary the range of the fit. This 1\% shift is negligible compared to the error on the
BAO peak point we obtain (as discussed in Section \ref{sec:decals_and_darksky_measurements}) 
and thus we believe it is subdominant. The constraints on the BAO peak $s_m$ 
for the wedge correlation function is obtained after fully marginalising all other parameters 
in $x_p$. We adopt a standard likelihood, 
$\mathscr{L} \propto \mathrm{exp}(-\chi^{2}/2)$ where the function $\chi^{2}$ is defined as,
\begin{align}
    \chi^2_{\mu_i}(x_p)= \sum_{s,s'}(\xi_{\rm mod}(s)-\xi_{w}(s,\mu_i))C^{-1}_{s;s'}(\mu_i)
                          \notag\\(\xi_{\rm mod}(s')-\xi_{w}(s,\mu_i))
\end{align}
where $\xi_{\rm mod}(s)$ is the model correlation function as given by Eq. \ref{eqn:empirical_fit}, 
$\xi_{w}(s,\mu_i)$ is the observed correlation function for the $i^{\rm th}$ $\bar{\mu}$ bin and 
$C^{-1}$ is the inverse covariance matrix.

\subsection{Covariance matrix from EZmock samples}\label{sec:covariance_matrix}

A random catalogue that is approximately 20 times the mock data is provided for the EZmock samples, but 
contains only the angular component (RA,DEC). Since the density of the DECaLS randoms is only 5 times
the data catalogue, we dilute the EZmock randoms to the same density to ensure consistency in our 
results. For the redshift component, we follow the same method of randomly extracting $\sigma_{z}$ 
from the data catalogue.

\begin{figure}
    \centering
	\includegraphics[width=0.47\textwidth]{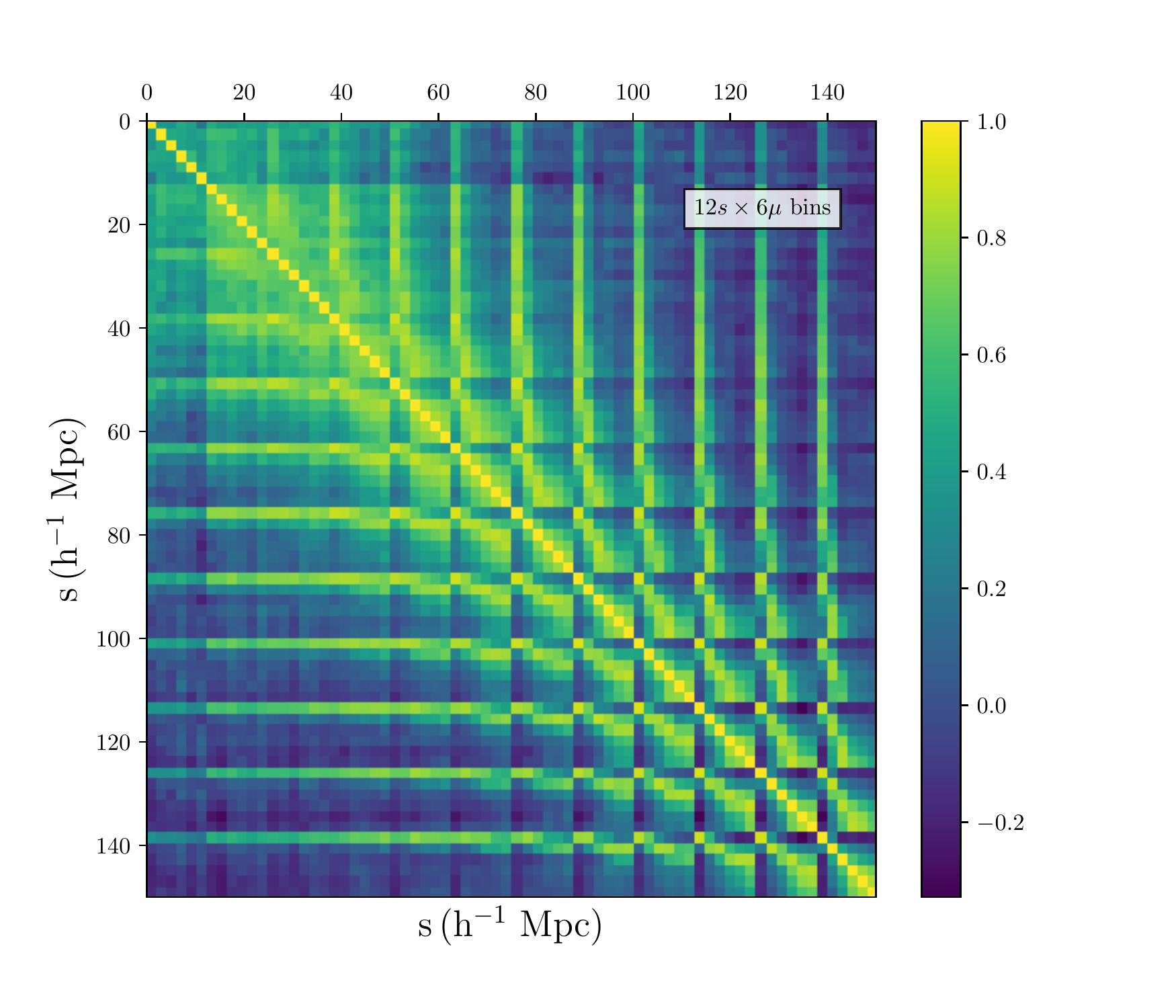}  
	\caption{The correlation matrix for the $0.6<z_{phot}<0.8$ sample using 12 bins in $s$          
	         and 6 bins in $\mu$ computed using Eq. \ref{eqn:corrmatrix}.} 
	\label{fig:corrmat}   
\end{figure}

We calculate the covariance matrix which is given by, 
\begin{equation}\label{eqn:cov_matrix}
C^{ij}_{w}(\xi^i_{w},\xi^j_{w})=\frac{1}{N-1}\sum^{N}_{n=1}[\xi^n_{w}(\vec{x}_i)-\overline{\xi}_{w}(\vec{x}_i)][\xi^n_{w}(\vec{x}_j)-\overline{\xi}_{w}(\vec{x}_j)],
\end{equation}
where the total number of simulations is given by $N$. 
The $\xi^n_{w}(\vec{x}_i)$ represents the value of the wedge correlation 
function of $i^{th}$ bin of $\vec{x}_i$ in the $n^{th}$ realisation, and 
$\overline{\xi}_{w}(\vec{x}_i)$ is the mean value of $\xi^n_{w}(\vec{x}_i)$ over all the 
realisations. Due to the limited number of mock samples (100) that we have, we use 12 bins in $s$. 
We obtain the correlation matrix as, 
\begin{equation}\label{eqn:corrmatrix}
    C_{ij} = \frac{\mathrm{Cov}(\xi_{i},\xi_{j})}{\sqrt{\mathrm{Cov}(\xi_{i},\xi_{i})\mathrm{Cov}(\xi_{j},\xi_{j})}}
\end{equation}
which is plotted in Figure \ref{fig:corrmat}. 

The number of realisations exceeds the number of $(s,\mu)$ bins of 72 bins 
($12(s $bins)$\times 6(\mu $bins)), 
and the inverse of $C_{ij}$ is well defined and thus does not require any de-noising 
procedures such as singular value decomposition. 
Additionally, we also count the offset caused by the finite number of realisation
\citep{Hartlap_2007} as, 
\begin{equation}\label{eqn:percival}
C^{-1}=\frac{N_{mocks} - N_{bins}-2}{N_{mocks}-1}\ \hat{C}^{-1}\ ,
\end{equation}
where $N_{bins}$ denotes the total number of $i$ bins.
As mentioned in Section \ref{sec:fitting_procedure}, we fit the correlation function within the 
range $30.0<s(h^{-1}\mathrm{Mpc})<130.0$. Thus, the number of $s$ bins in this range is 9. As the 
final 3 $\mu$ bins along the LOS do not contain any BAO information, we perform the fitting by 
ignoring them. Thus, the shape of the matrix is 27$\times$27 and for 100 mock realisations, the 
factor in Eq. \ref{eqn:percival} becomes 0.71. 
Apart from the above correction factor, an additional correction to the inverse 
covariance matrix is proposed by \cite{Percival_2014}, which is given by, 
\begin{equation}
    m_{1} = \frac{1+B(n_{b}-n_{p})}{1+A+B(n_{p}+1)}
\end{equation}
where $n_{b}$ is the number of bins used for the two-point correlation measurements,
$n_{p}$ is the number of parameters measured and the $A$ and $B$ terms are given by, 
\begin{subequations}
\begin{equation}
    A = \frac{2}{(n_{s}-n_{b}-1)(n_{s}-n_{b}-4)}
\end{equation}

\begin{equation}
    B = \frac{(n_{s}-n_{b}-2)}{(n_{s}-n_{b}-1)(n_{s}-n_{b}-4)}
\end{equation}
\end{subequations}
where $n_{s}$ is the number of simulations used for the covariance matrix calculations. 
Applying the square root of this expression to the measured standard deviation
should take care of the extra correction. For our binning scheme as mentioned above (with 100 mock 
realisations), we get $\sqrt{m_1}=0.89$ which is significantly smaller than the correction factor 
we already apply. 
The BAO peak position $s_m$ of the wedge correlation function is found by fitting the 
phenomenological model by considering full covariance using $C_{ij}$.

\begin{figure*}
\includegraphics[width=0.98\textwidth]{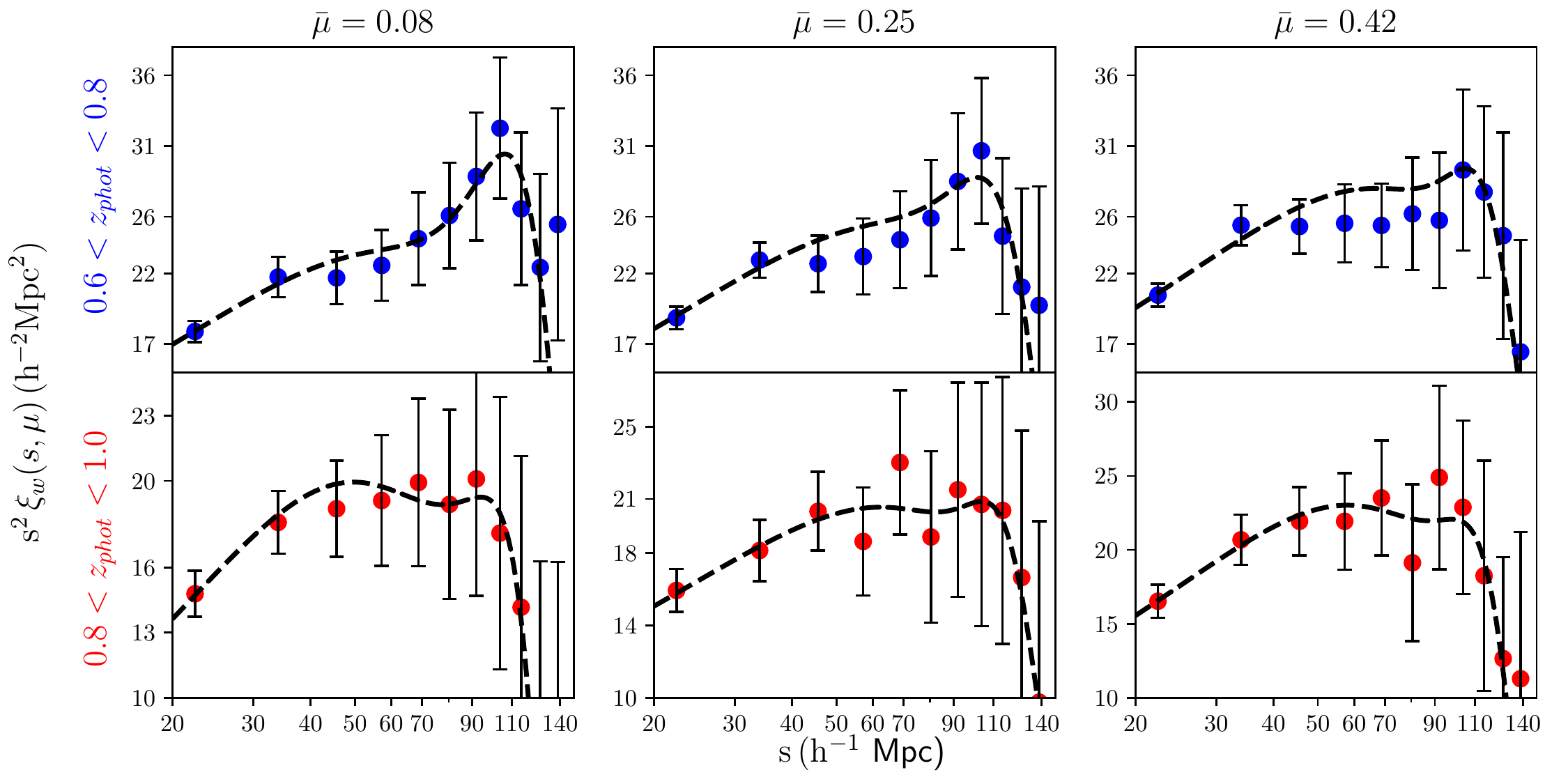}   
\caption{The correlation function \xismu (multiplied by $s^{2}$) 
	     calculated by splitting into wedges of $\mu$ for the $0.6<z_{phot}<0.8$ sample 
	     (given by blue dots) and for the $0.8<z_{phot}<1.0$ sample (given by red dots).
	     The first, second and third columns in the figure represent $\bar \mu=0.08$, 
	     0.25 and 0.42 bins respectively. The dashed black lines in each plot 
	     show the best-fit (maximum likelihood) 
	     obtained from the empirical model by applying the MCMC technique.
	     The error bars plotted are the square root of the diagonal elements of the full 
	     covariance matrix as mentioned in Eq. \ref{eqn:cov_matrix}.}
\label{fig:xismu}   
\end{figure*}

\begin{figure*}
    \centering
	\includegraphics[width=0.48\textwidth]{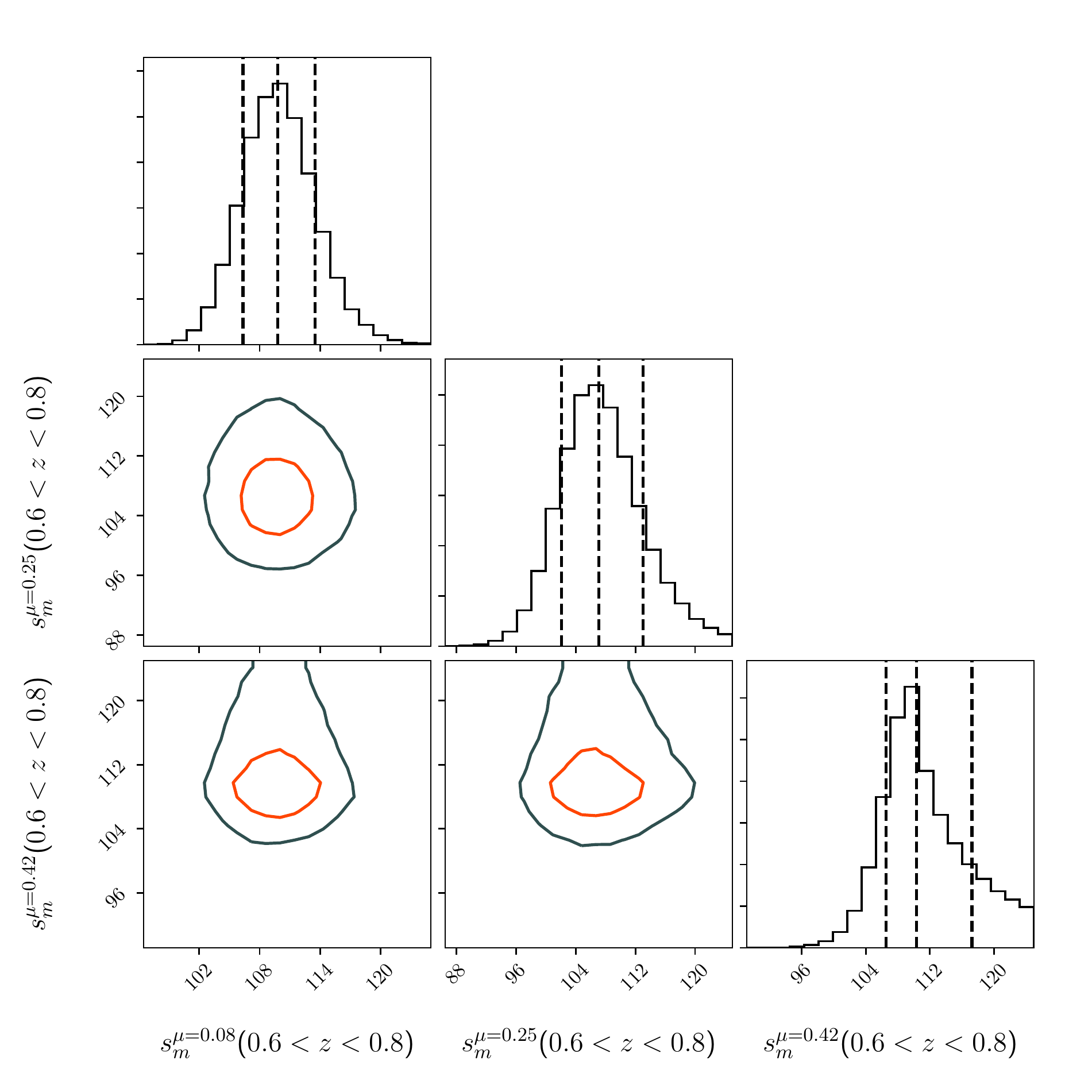}
	\includegraphics[width=0.48\textwidth]{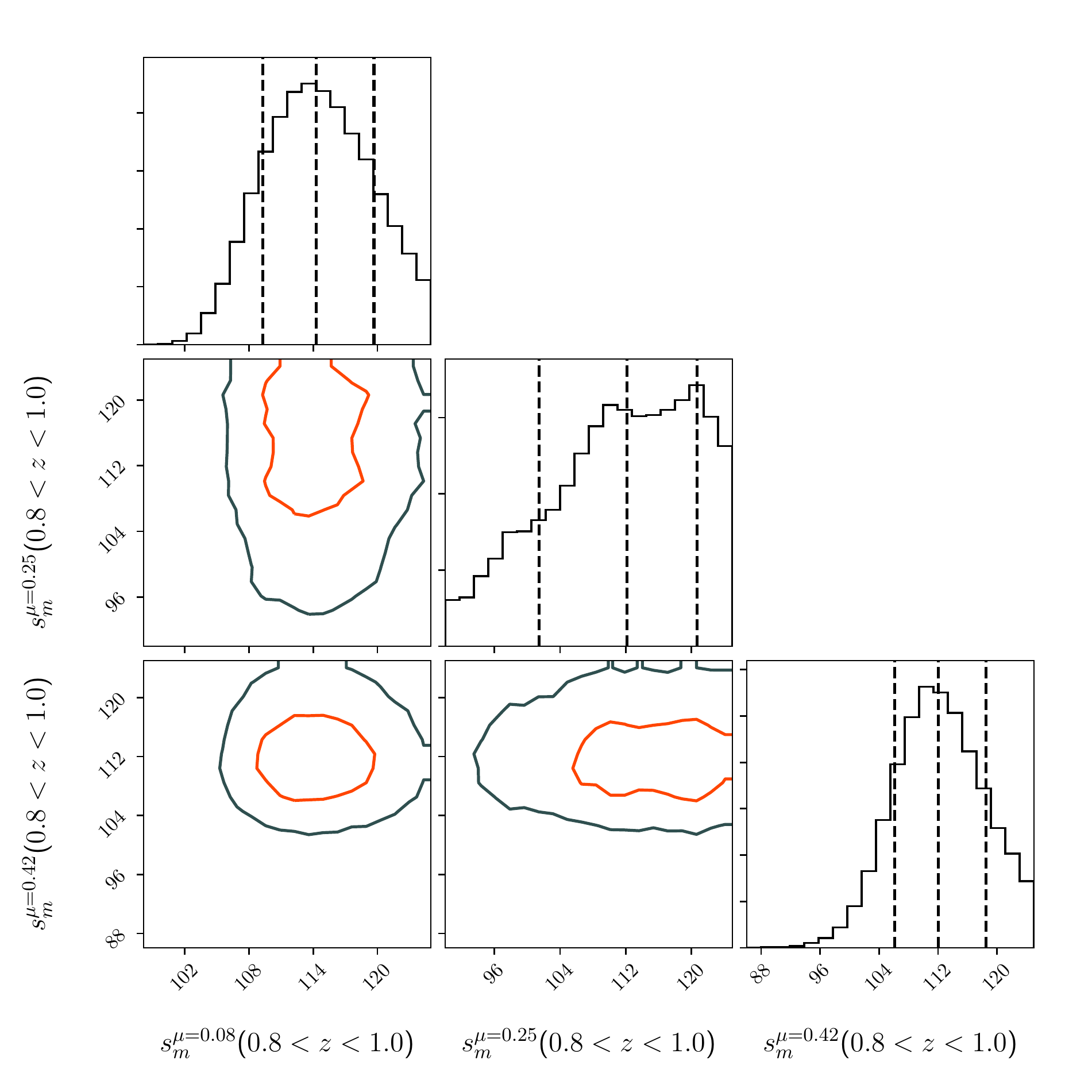}
	\caption{\textit{Left panel:}
	         The marginalised posterior distribution of the peak point $s_{m}$ obtained from the 
	         MCMC analysis using Equation \ref{eqn:empirical_fit} for all the 3 $\mu$ bins in the 
	         redshift range $0.6<z_{phot}<0.8$. 
	         The covariances between the different $s_{m}$ values is shown in the 
	         contour plots and the orange circle encompasses all points within the 1$\sigma$ region.
	         The marginalised distribution for each $s_{m}$ independently is shown 
	         in the histograms along the diagonal.
	         \textit{Right panel:} The same for the $0.8<z_{phot}<1.0$ sample.} 
	\label{fig:sm_mcmc_posterior}   
\end{figure*}

\subsection{DECaLS and Dark Sky mock acoustic-scale measurements from wedge correlation function}
\label{sec:decals_and_darksky_measurements}

From the parent DECaLS DR8 sample within the redshift range $0.3<z_{phot}<1.2$, 
we choose two redshift cuts between $0.6<z_{phot}<0.8$ and $0.8<z_{phot}<1.0$ for our analysis. 
The reason for choosing these two redshift ranges is because the redshift distribution of the sample
peaks at $z_{phot} \approx 0.7$ as it can be seen from Figure \ref{fig:RA_vs_DEC}. Thus, we expect 
to have the maximum number of galaxies around this redshift range. 
The redshift uncertainty scales with redshift, so we quote the mean values for our two redshift 
cut samples in Table \ref{tab:table1}.
The wedge correlation functions are calculated using Eq. \ref{eqn:wed_xismu} by using 6 $\mu$ bins
of thickness $\Delta\mu = 0.17$. 
The $\xi_w(s,\mu_i)$ calculated from the first three $\mu$ bins for the two 
redshift samples is shown in Figure \ref{fig:xismu}. 
We fit \xismu using Eq. \ref{eqn:empirical_fit} by following the MCMC procedure described in 
Section \ref{sec:fitting_procedure} and the values of the BAO peak $s_{m}$ from the fit 
is provided in Table \ref{tab:table2}. 

One can observe that for both the redshift
samples, the BAO signal is diluted as $\mu$ increases, and is also more clearly visible for the 
first redshift sample compared to the second redshift sample. This is due to the fact that the 
photometric redshift errors scale proportional to $\sigma_{0}\times(1+z)$, and thus 
the isotropy along the LOS is destroyed more strongly for the high redshift sample compared to the 
low redshift one. It can also be seen from Table \ref{tab:table2} that for the same reason, 
the BAO peak appears at greater $s$ at greater $\mu$ for both the redshift samples. This trend is
more strongly observed for the first redshift sample with $s_{m}$ increasing from 109.7
h$^{-1}$Mpc to 111.0 h$^{-1}$Mpc from $\bar{\mu}=0.08$ to 0.42. 
The errors on $s_{m}$ gradually increase with increasing $\bar{\mu}$ as expected. 

For both the redshift ranges and all the 3 $\bar{\mu}$ bins used, we fit the correlation 
function within the range $30.0<s(h^{-1}\mathrm{Mpc})<130.0$. When using log binning instead of 
linear binning we note that the BAO peak results slightly shift to higher values. However, the effect
is of the order of 1\%, which is well below the estimated accuracy of the BAO peak position,
which is between 4-10\% for our samples. We use the $\chi^{2}/dof$ goodness of fit indicator to 
validate the performance of our empirical model to fit the correlation function. 
The overall fit to the $0.6<z_{phot}<0.8$ sample yields a $\chi^{2}/dof$ = 12/9
,including all cross-covariance between $\bar\mu$ bins. We obtain a $\chi^{2}/dof$ = 15/9 for the 
$0.8<z_{phot}<1.0$ sample.

It has been shown from previous studies \citep{Ross_2012,Ross_2017} that data from the different 
$\mu$ bins is expected to be correlated and that the results from splitting the clustering 
by $\mu$ show a slight decrease in the BAO information content with increasing $\mu$. Thus, we 
perform the fit using the full data vector including all the $\mu$ bins. By performing the fit using 
the full covariance matrix, we make sure that
the correlations between the different $\mu$ bins that exist are taken into account. 
The one and two dimensional 
projections of the posterior probability distribution of the $s_{m}$ parameter from the MCMC chains 
for the two redshift samples is shown in 
Figure \ref{fig:sm_mcmc_posterior}. The marginalized distribution for each $s_{m}$ value from each 
$\bar \mu$ bin is shown independently in the histograms along the diagonal and the marginalized two 
dimensional distributions in the other panels. We find that the correlation 
between the $s_{m}$ values obtained at the different $\mu$ bins for both the redshift samples to be 
minimal.

\begin{table}
\begin{center}
\begin{tabular}{c c c } 
\hline\hline
Redshift range & $\bar{\mu}$ bin & $s_{m}$ (h$^{-1}$Mpc)  \\ [0.5ex] 
\hline
                   & 0.08 & $109.7^{+3.6}_{-3.4}$ \\ 
$0.6<z_{phot}<0.8$ & 0.25 & $107.1^{+5.8}_{-5.0}$ \\
                   & 0.42 & $111.0^{+7.8}_{-4.3}$ \\
\hline
                   & 0.08 & $111.2^{+5.3}_{-4.9}$ \\ 
$0.8<z_{phot}<1.0$ & 0.25 & $111.4^{+9.1}_{-9.4}$ \\
                   & 0.42 & $112.0^{+6.4}_{-5.8}$ \\
\hline
\hline
\end{tabular}
\caption{Results of fitting the correlation function (plotted using the dotted lines in Figure 
         \ref{fig:xismu}) for the two redshift samples and in the 3 
         $\bar{\mu}$ bins using Eq. \ref{eqn:empirical_fit}. The $s_{m}$ is the BAO peak point 
         obtained from the fit and the units are in h$^{-1}$Mpc.}
\label{tab:table2}
\end{center}
\end{table}

\begin{figure*}
\includegraphics[width=0.98\textwidth]{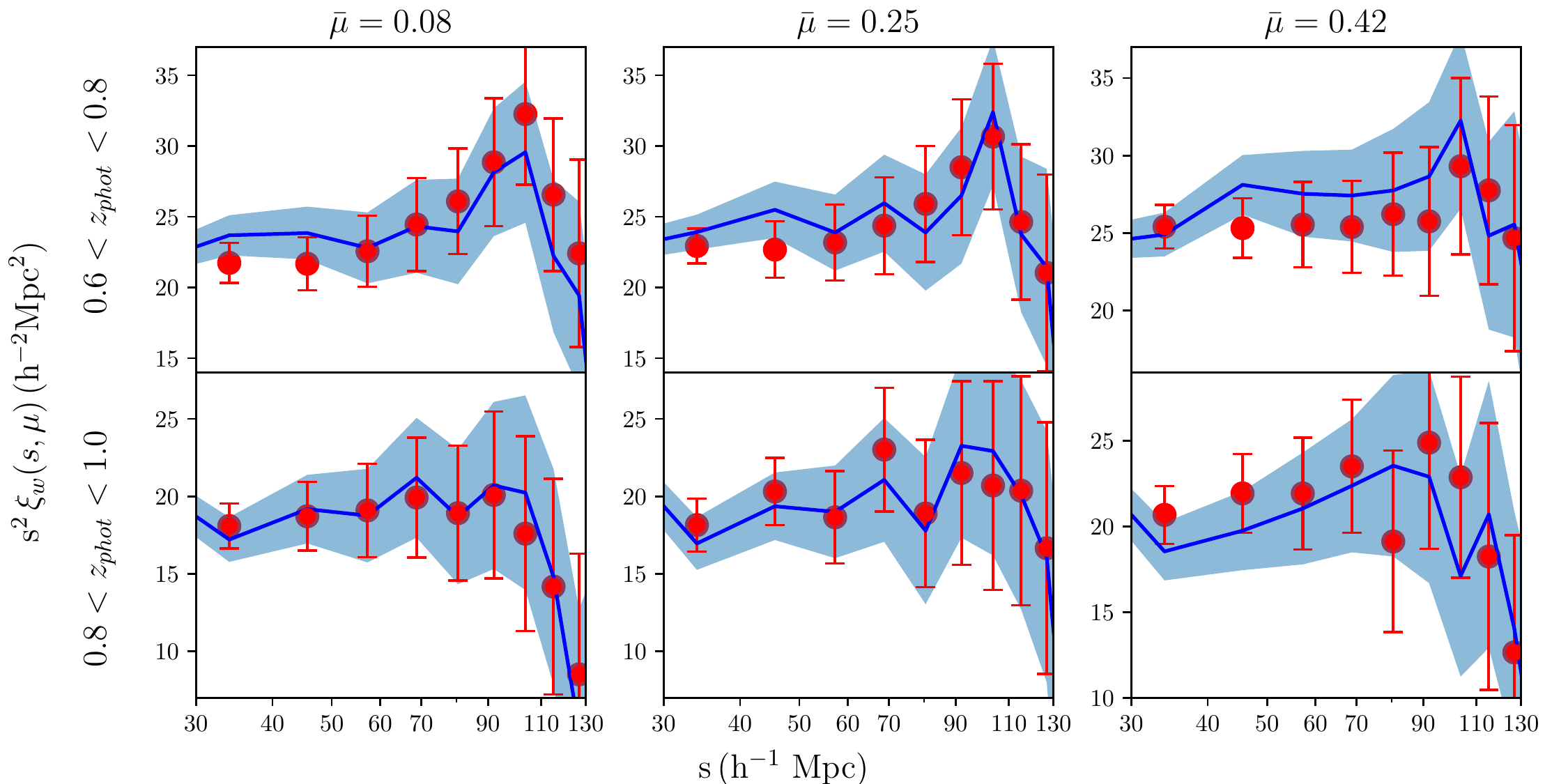}
\caption{\textit{Top panel:} \xismu (multiplied by $s^{2}$) calculated for the DECaLS data (dotted 
         points in red, same as in Figure
         \ref{fig:xismu}) compared with \xismu calculated for the Dark Sky simulation (solid lines 
         in blue) within the redshift range $0.6<z_{phot}<0.8$ for the three $\mu=0.08,0.25,0.42$ 
         bins from left to right. 
         \textit{Bottom panel:} Same plot for the samples within the redshift range $0.8<z_{phot}<1.0$.
         A constant value of 0.0005 and 0.0010 has been added to the \xismu from the Dark Sky mocks 
         for the $0.6<z_{phot}<0.8$ and $0.8<z_{phot}<1.0$ samples respectively so that the amplitudes
         match with the \xismu from the DECaLS data. 
         The error bars are the square root of the diagonal elements of the full 
	     covariance matrix as mentioned in Eq. \ref{eqn:cov_matrix}. The error bars for the Dark Sky 
	     sample are represented by the shaded blue region.}
\label{fig:xi_decals_vs_darksky}
\end{figure*}

For validating our clustering results obtained on the DECaLS data, 
we use the LRGs from 
the Dark Sky mock catalogue. For mimicking the photometric redshifts from the 
DECaLS data, we use the same procedure followed for generating photo-z's for our EZmock sample. 
We obtain $\sigma_{z}$'s randomly from the parent DECaLS sample, but by restricting to galaxies 
of similar redshifts. To compare the correlation function results obtained from the two 
redshift samples of the DECaLS catalogue, we compute \xismu from the Dark Sky photometric redshift 
catalogues with the same redshift cuts and use the same binning scheme. We find that by using the 
true values of \xismu for the two photometric redshift samples from the Dark Sky mocks, 
the amplitudes of \xismu
do not match with the \xismu from the DECaLS data. However, by adding a constant value of 0.0005 and 
0.0010 to the \xismu obtained from the Dark Sky mocks for the $0.6<z_{phot}<0.8$ and $0.8<z_{phot}<1.0$ 
samples, we see that the amplitudes match well. The results are presented in Figure 
\ref{fig:xi_decals_vs_darksky}. 

There are two key features that we are 
interested in when comparing the data and the mock catalogue. One is the amplitude of 
\xismu and the other is the location of the BAO peak. We find that the amplitude of the 
clustering measurements from the Dark Sky mock catalogue (given by solid blue line) match with 
the amplitude of the clustering measurements from the DECaLS sample (given by the red scatter points)
for both the redshift samples in all the 3 $\bar{\mu}$ bins after adding the constant values to our 
redshift samples as described above. To statistically compare the linear 
correlation between the two samples, we use the non-parametric two-sample Kolmogorov-Smirnov test 
(KS test). The null hypotheses for the KS test is that the distributions are the same and to reject the 
null hypotheses we require a $p$-value less than 0.05. For the $0.6<z_{phot}<0.8$ sample, in all the 
three $\bar{\mu}$ bins, the minimum $p$-value we obtain is 0.45. For the $0.8<z_{phot}<1.0$ sample, 
in all the three $\bar{\mu}$ bins,
the minimum $p$-value we obtain is 0.48. These results show that the \xismu obtained from the Dark Sky 
and the DECaLS samples are similar. 
We repeat the MCMC procedure to obtain the BAO peak for the Dark Sky sample and find that the 
location of the BAO peak from the Dark Sky samples for both the redshift ranges agree with the 
DECaLS sample, at least within 1$\sigma$. A similar result has been obtained by
\cite{Ross_2017} by doing a comparison between mock samples and model curves using mock 
photometric data.

\begin{figure}
	\includegraphics[width=0.47\textwidth]{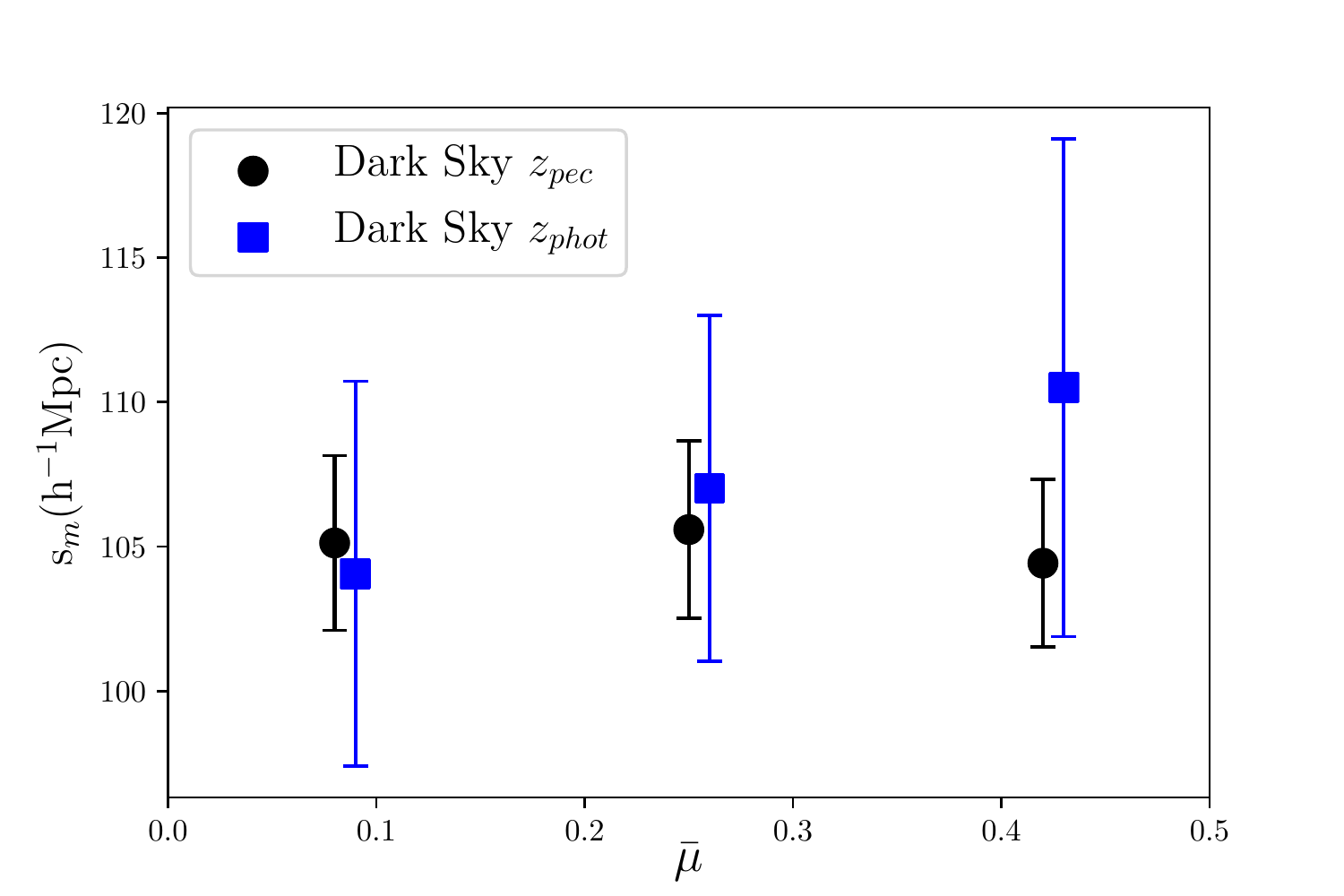}  
	\caption{The x-axis denotes the 3 $\bar \mu$ bins we have used and the y-axis denotes the 
		     value of $s_{m}$ obtained from the empirical fit for the Dark Sky $z_{pec}$ sample
		     (black dots) and the Dark Sky $z_{phot}$ sample (blue dots) within the redshift range 
		     $0.6<z_{phot}<0.8$. The $\bar \mu$ bins for the $z_{phot}$ sample have been shifted 
		     by 0.01 for better visualisation.} 
	\label{fig:mu_vs_sm_darksky_vs_darksky}   
\end{figure}

We also verify the internal consistency of the BAO peaks obtained from the Dark Sky photometric 
catalogue by comparing it with the BAO peaks obtained from the true redshift ($z_{pec}$, 
cosmological redshift with peculiar velocity added) catalogue for the same $\bar \mu$ bins. 
The wedge correlation functions from the 
three $\bar \mu$ bins are calculated using Eq. \ref{eqn:wed_xismu} and the $s_{m}$ values
obtained from the two samples are plotted in Figure \ref{fig:mu_vs_sm_darksky_vs_darksky}. For 
all the three $\bar \mu$ bins, it can be seen that the $s_{m}$ values from the photometric samples
are within 1$\sigma$ compared to the $s_{m}$ values from the $z_{pec}$ sample, with the 
$1\sigma$ errors on $s_{m}$ being larger for the $z_{phot}$ sample.

\section{Measured cosmic distances}\label{sec:measured_cosmic_distance}

In the previous section, we measured the BAO peak position $s_{m}$ for our DECaLS sample at 
different $\bar\mu$
bins. In this section, we explain the theoretical model which we use to obtain the 
theoretical correlation function $\xi_{\rm th}(s,\mu)$. The $\xi_{\rm th}$ is a function of both 
$s$ and $\mu$, which can then be used to translate our measured BAO peak positions to 
physical distances. 

The volume distance at a redshift is given by,
\begin{equation}
    D_{V}(z) = \left[(1+z)^{2}D_{A}(z)^{2}cz/H(z)\right]^{1/3}
\end{equation}
and is measured through the BAO by exploiting the monopole correlation function ($\xi_{0}$). 
It has been shown from 
previous studies \citep{Estrada_2009,Srivatsan_2019} that for photometric redshift samples, 
the BAO peak is smeared out in $\xi_{0}$. Thus, using the wedge approach, both the transverse and 
radial cosmic distances can be separately measured.

\subsection{Theoretical model for the correlation function}

We compute the theoretical correlation function $\xi_{\rm th}(s,\mu)$ in redshift space exploiting 
the improved power spectrum based upon the TNS model as,
\ba\label{eq:xi_eq}
\xi_{\rm th}(s,\mu)&=&\int \frac{d^3k}{(2\pi)^3} \tilde{P}(k,\mu')e^{i{\bf k}\cdot{\bf s}}\nn\\
&=&\sum_{\ell:{\rm even}}\xi_\ell(s) {\cal P}_\ell(\mu)\,,
\ea
with ${\cal P}$ being the Legendre polynomials. Here, we define $\mu=\pi/s$ and 
$s=(\sigma^2+\pi^2)^{1/2}$. The moments of the correlation function, $\xi_\ell(s)$, are defined by,
\ba
\xi_\ell(s)=i^\ell\int\frac{k^2dk}{2\pi^2}\,\tilde{P}_\ell(k)\,j_\ell(ks)\,.
\ea
The multipole power spectra $\tilde{P}_\ell(k)$ are explicitly given by,
\ba
\tilde P_0(k)&=&p_0(k),\nn\\
\tilde P_2(k)&=&\frac{5}{2}\left[3p_1(k)-p_0(k) \right],\nn\\
\tilde P_4(k)&=&\frac{9}{8}\left[35p_2(k)-30p_1(k)+3p_0(k)\right],\nn\\
\ea
where we define the function $p_m(k)$: 
\ba
p_m(k)&=&\frac{1}{2}\sum_{n=0}^4\frac{\gamma(m+n+1/2,\kappa)}{\kappa^{m+n+1/2}}\,Q_{2n}(k)
\ea
with $\kappa=k^2\sigma_p^2$. The function $\gamma$ is the incomplete gamma function of the 
first kind: 
\ba
\gamma(n,\kappa)=\int^{\kappa}_0dt\, t^{n-1}\,e^{-t}\,.
\ea
The $Q_{2n}$ is explained below.

The observed power spectrum in redshift space $\tilde{P}(k,\mu)$ is written in the following 
form;
\ba
\label{eq:pkred_in_Q}
\tilde{P}(k,\mu) =\sum_{n=0}^8\,Q_{2n}(k)\mu^{2n}\,G^{\rm FoG}(k\mu\sigma_p)\,,
\ea
where the velocity dispersion $\sigma_p$ is set to be a free parameter for FoG effect, 
and the function $Q_{2n}$ are given by,
\ba\label{eq:Q}
Q_0(k)&=&P_{\delta\delta}(k) ,\nn\\
Q_2(k)&=&2P_{\delta\Theta}(k) + C_2(k),\nn\\
Q_4(k)&=&P_{\Theta\Theta}(k) + C_4(k),
\ea
where $C_n$ includes the higher order polynomials caused by the correlation between density and 
velocity fluctuations, and $P_{XY}(k)$ denotes the 
power spectrum in real space. The standard perturbation model exhibits the ill-behaved 
expansion leading to the bad UV behaviour. In this manuscript, we use the resummed perturbation 
theory {\tt RegPT} which is regularised by introducing UV cut-off~\cite{Taruya:2012ut}.
The auto and cross spectra of $P_{XY}(k)$ are computed up 
to first order, and higher order polynomials are computed up to zeroth order, 
which are consistent in the perturbative order.

\begin{figure*}
	\includegraphics[width=0.49\textwidth]{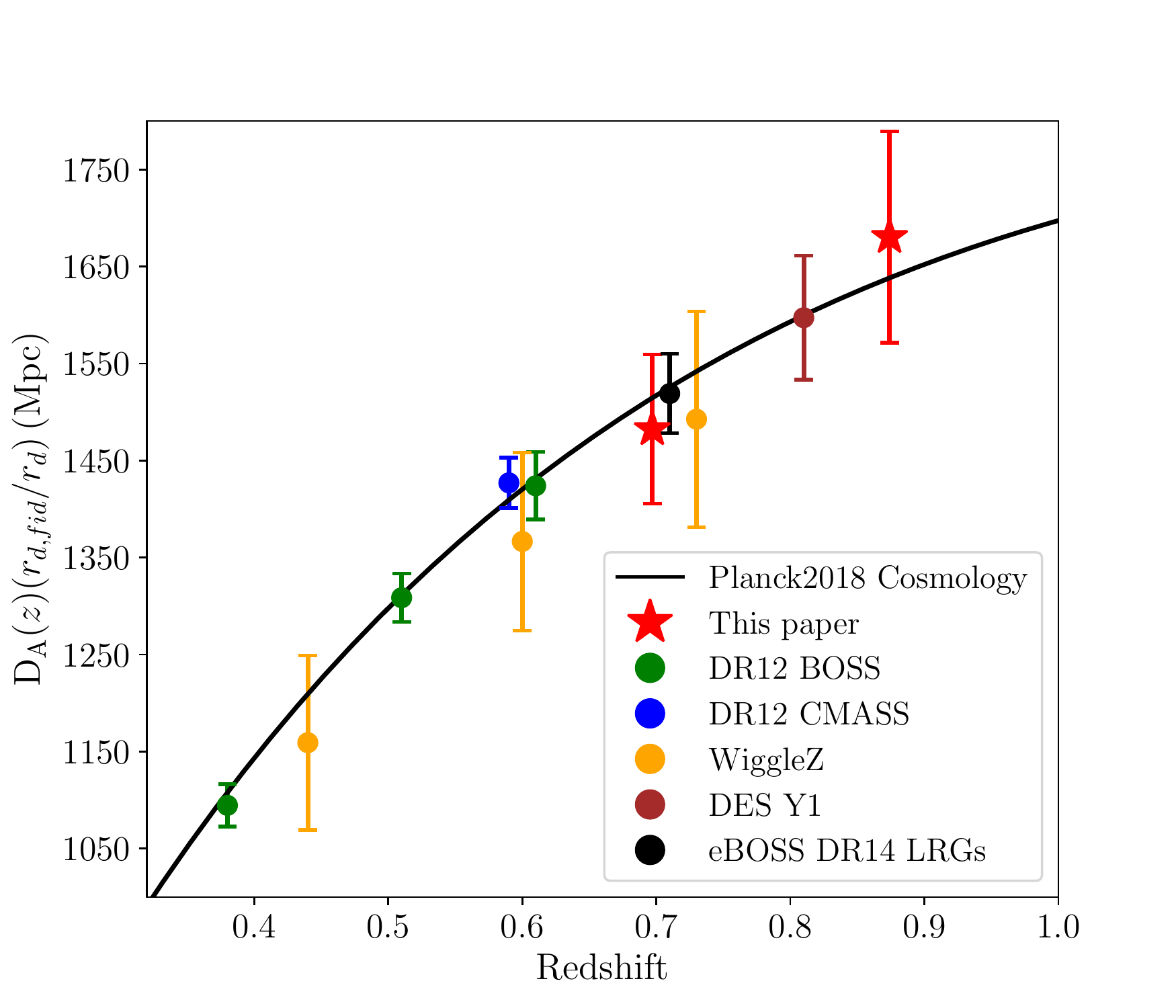} 
	\includegraphics[width=0.49\textwidth]{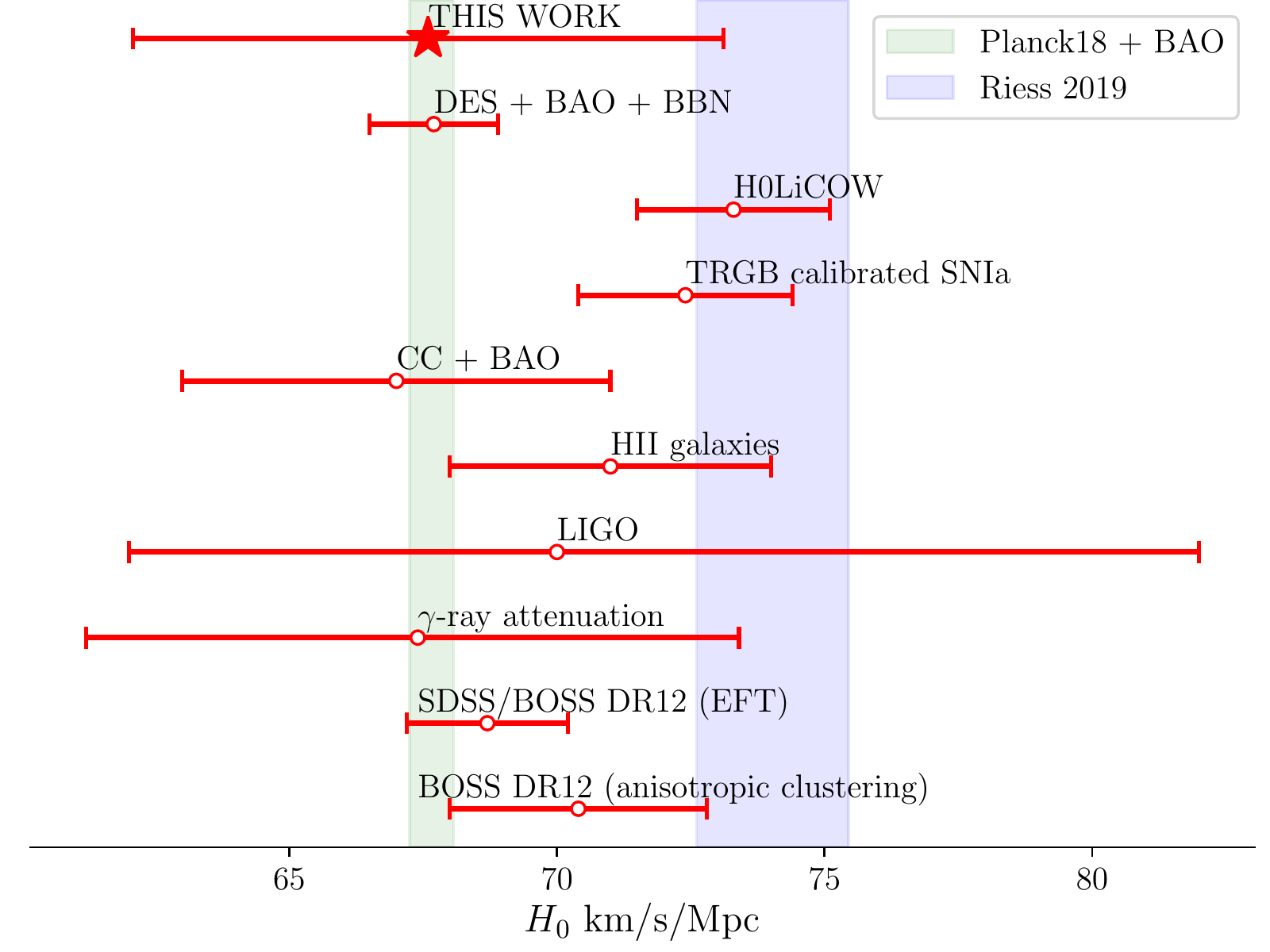} 
	\caption{\textit{Left panel:} The $D_{A}(z)\times(r_{d,fid}/r_{d})$ measurements obtained from 
	         the two redshift samples
		     are plotted in red along with the values of $D_{A}(z)\times(r_{d,fid}/r_{d})$ measured 
		     by other surveys 
		     (as colour coded in the legend and mentioned in the text). 
		     The solid black line corresponds to 
		     the theoretical predictions for $D_{A}(z)$ as a function of redshift obtained
		     using the cosmological parameters measured from \cite{Planck_2018}. 
		     Units are in Mpc. The normalised likelihood $\mathscr{L}$ = exp$(-\Delta\chi^{2}/2)$ 
		     for $D_{A}(z)$ obtained 
		     for the $0.6<z_{phot}<0.8$ (solid red line) and the $0.8<z_{phot}<1.0$ sample (dotted 
		     red line) is plotted in the inset plot. 
		     \textit{Right panel:} The value of $H_{0}$ obtained using the 
		     likelihoods from our $D_{A}$ measurements (on top) compared with $H_{0}$ measurements
		     from different probes with $1\sigma$ error bars. The $1\sigma$ error from the 
		     \cite{Planck_2018} (constraints including BAO) is plotted in light green and the 
		     $1\sigma$ error from \cite{Riess_2019} is plotted in light blue. 
		     From bottom to top, enumerated on 
		     the vertical axis, \cite{bossdr12_h0} (BOSS DR12 constraints from anisotropic clustering
		     measurements), \cite{sdssbossdr12_h0} (SDSS/BOSS DR12 constraints using effective field
		     theory), 
		     \cite{gammabao_h0} ($\gamma$-ray attenuation), 
		     \cite{ligo_h0} (LIGO binary black hole merger GW170817), 
		     \cite{hii_h0} (HII galaxies), \cite{ccbao_h0} (cosmic chronometers + BAO),
		     \cite{trgb_h0} (TRGB calibrated SNIa), \cite{holicow_h0} 
		     (H0LiCOW, gravitationally lensed quasars), \cite{desbaobbn_h0} 
		     (DES clustering + weak lensing).}
	\label{fig:z_vs_DA}   
\end{figure*}

Although cosmic distances are estimated using the BAO at linear regimes, there are smearing effects
at small scales which need to be computed. These small scale corrections are included to make the final
precise constraints on the BAO \citep{Taruya:2010mx}. We make use of an improved model of the 
redshift-space power spectrum \citep{Taruya:2010mx}, in which the coupling between the density and 
velocity fields associated with the Kaiser and the FoG effects is perturbatively incorporated 
into the power spectrum expression. The resultant includes nonlinear corrections consisting of 
higher-order polynomials~\citep{Taruya:2010mx}:
\ba
\hat{P}(k,\mu) &=& \big\{P_{\delta\delta}(k) + 2\mu^2 P_{\delta\Theta}(k) + \mu^4 
P_{\Theta\Theta}(k) \nn\\
&+& A(k,\mu) + B(k,\mu)\big\}G^{\rm FoG}
\label{eq:TNS10}
\ea
Here the $A(k,\mu)$ and $B(k,\mu)$ terms are the nonlinear corrections, and are expanded as 
power series of $\mu$. Those spectra are computed using the fiducial cosmological parameters. 
The FoG effect $G^{\rm FoG}$ is given by the simple Gaussian function which is written as,
\ba
G^{\rm FoG} \equiv  \exp{\left[-(k\mu\sigma_p)^2\right]}
\ea
where $\sigma_p$ denotes one dimensional velocity dispersion. 

Thus the theoretical correlation 
function $\xi_{\rm th}(s,\mu)$ is parameterised by, 
\begin{equation}\label{eqn:tns_params}
    x_{\rm th} = (D_{A},H^{-1},G_{b},G_{\Theta},\sigma_{p})
\end{equation}
wherein $G_{b}$ and $G_{\Theta}$ are the normalised density and coherent motion growth functions.
When working with photo-z samples, the effect of 
the photo-z error on the correlation function is incoherent. Thus, an extra parameter is needed 
for the theoretical template to model $\xi_{\rm th}(s,\mu)$ as a function of the photo-z error, 
but it is not well understood. So, we use Eq. \ref{eqn:empirical_fit} instead to fit our observed \xismu.
This functional form only assumes a power-law at small scales and a Gaussian function to fit 
the BAO peak at large scales and seems to model \xismu quite well as we can see from Fig.
\ref{fig:xismu}.  

In our previous work \citep{Srivatsan_2019} we have verified that the BAO feature from the theoretical 
correlation function is weakly dependent on the growth functions and $\sigma_{p}$. 
We have verified that changing the value of $\sigma_{p}$ by $\pm$10\% does alter the location of the 
BAO peak, but only by less than 0.1\% for samples at $\bar \mu$ close to 0 and less than 1\% for 
samples at $\bar \mu$ close to 1, which is negligible.
Thus, when we fit the cosmic distances, 
we fix $\sigma_{p}$. To find the best-fit $\sigma_{p}$, we vary it within 
$3.0<\sigma_{p}(\mathrm{h}^{-1}\mathrm{Mpc})<6.0$ (by fixing $D_{A}$ and $H^{-1}$ to their fiducial 
values), compute the theoretical correlation function $\xi_{\rm th}(s,\mu)$ and use the $\sigma_{p}$ for 
which $\xi_{\rm th}(s,\mu)$ matches best with our measured \xismu. 
The best fit $\sigma_p$ used in this works is $\sigma_{p}=4.8\,\mathrm{h}^{-1}\mathrm{Mpc}$, and we 
fix it for both the redshift ranges. 
 
Note that we apply the TNS model for computing the theoretical BAO peaks to fit the measured 
data. It has been shown in \cite{Srivatsan_2019} that using a simple coordinate transformation to 
transform the theoretical BAO peaks from one cosmology to another results in incorrect values of the 
BAO peak especially at high $D_{A}$ and $H^{-1}$ values. Thus, the TNS model is adopted to determine the 
theoretical BAO points rather than a simple coordinate transformation.

\subsection{Cosmic distance measurements}\label{sec:dist_measures_final}

In Section \ref{sec:decals_and_darksky_measurements}, we obtained the $s_{m}$ values for our two redshift
samples from the DECaLS data. In this section, we explain how we use a two-step process to go from 
$s_{m}$ to measured cosmic distances $D_{A}$ and $H^{-1}$. 
As mentioned in the previous section, $\xi_{\rm th}$ is a function of $s$ and $\mu$ and depends on 
5 parameters as mentioned in Eq. \ref{eqn:tns_params}. Since $\xi_{\rm th}(s,\mu)$ 
is weakly dependent on the growth functions ($G_{b}$ and $G_{\Theta}$) and $\sigma_{p}$, 
we vary the tangential and radial 
distance measures from the fiducial values of $D_{A}$ and $H^{-1}$ for the two redshift samples when 
we fit the cosmic distances. We use a $13\times13$ grid for varying $D_{A}$ and $H^{-1}$ and 
both $D_{A}$ and $H^{-1}$ are sampled within 
$0.6( param^{fid}) <  param^{fid} < 1.4( param^{fid})$, where $param$ is either $D_{A}$
or $H^{-1}$. For each of these parameter set we obtain a $\xi_{\rm th}(s,\mu)$ for the given $\mu$. 
We then fit each of our $\xi_{\rm th}(s,\mu)$ function using Eq.\ref{eqn:empirical_fit} to obtain 
the theoretical $s_{m}$ values. 
We then compare the $\xi_{\rm th}(s,\mu)$ with our measured \xismu and compute the $\chi^{2}$ 
values using the $s_{m}$ values from the DECaLS data and 
the theoretical templates for the first 3 $\bar{\mu}$ bins (as we find that the BAO peak is 
washed-out for the last three $\bar{\mu}$ bins) taking into account the covariance
between the $s_{m}$ values between the different $\bar{\mu}$ bins that exist.

The uncertainty on the redshift determination prevents us from accessing the radial cosmic 
distance, and thus the BAO peak is not clearly visible for the $\xi(s,\bar \mu>0.5)$ correlation 
functions. Thus, we do not get tight constraints on $H^{-1}(z)$.
However, even after fully marginalising over $H^{-1}(z)$, the transverse 
cosmic distance $D_{A}(z)\times(r_{d,fid}/r_{d})$ is 
measured with good precision for both the redshift samples. The fiducial 
values of $D_{A}^{fid}$ for our cosmology at the two mean redshifts is
$D_{A}^{fid}(\bar z=0.697) = 1514$ Mpc and 
$D_{A}^{fid}(\bar z=0.874) = 1638$ Mpc and the measured values are:
\begin{equation}
    D_{A}(0.697) = (1499 \pm 77 \, \mathrm{Mpc})\left(\frac{r_{d}}{r_{d,fid}}\right),
\end{equation}
\begin{equation}
    D_{A}(0.874) = (1680 \pm 109 \, \mathrm{Mpc})\left(\frac{r_{d}}{r_{d,fid}}\right),
\end{equation}
These values correspond to distance measures of 5.14\% and 6.48\% precision
for the two redshift samples respectively. The 0.2\% statistical error on 
$r_{d}$ based on the \cite{Planck_2018} measurements only make a negligible 
contribution when added in the above equations. We compare our results with 
previous studies in Figure \ref{fig:z_vs_DA}. The constraints using four 
spectroscopic redshift surveys, i.e. \cite{WiggleZ_2011} (WiggleZ), 
\cite{Alam_2017} (DR12 BOSS), \cite{Chuang_2017} (DR12 CMASS)
and \cite{eBOSS_2018} (DR14 eBOSS) are plotted in yellow, green, blue and black
respectively. The constraints using the DES photometric redshift survey (\cite{DES_Y1}) is 
plotted in brown. The solid black line corresponds to the theoretical
predictions as a function of redshift obtained using the cosmological 
parameters from \cite{Planck_2018}.

The cosmological concordance model with the cosmological constant is assumed to 
be a cause of cosmic acceleration, with the Hubble constant unknown. 
The measured angular diameter distance at the two redshifts from the DECaLS 
sample and the prior information of the
sound horizon size and $w_{m} \equiv \Omega_{m}h^{2} = 0.1430\pm0.0011$ as 
determined by the Planck 
experiment (\cite{Planck_2018}) are used to get constraints on $H_{0}$. 
The $\chi^{2}$ values from $D_{A}(z)$ along with the $\chi^{2}$ values from 
$w_{m}$ are cumulatively summed up to get the final constraint on $H_{0}$.
We fit for the three parameters 
($\Omega_{m},\Omega_{m}h^{2},H_{0}$) using flat, wide priors which extend well
beyond the regions of high likelihood and have no effect on the 
cosmological fits and obtain a value of $H_{0}=67.59\pm5.52$ km/s/Mpc. 

We also make a comparison plot with $H_{0}$ measurements obtained from recent works using different
probes in the right panel of Figure \ref{fig:z_vs_DA}. 
Our $H_{0}$ value is measured with 8.1\% precision, whereas some of the estimates from other 
probes plotted in Figure \ref{fig:z_vs_DA} have a better precision. To quote a few, the 
HII galaxy data \citep{hii_h0} delivers a $\sigma_{H_{0}}/H_{0}$ = 4.9\%, the DES + BAO + BBN 
data delivers a $\sigma_{H_{0}}/H_{0}$ = 1.8\%. The reason for our conservative
estimate is because we only use the likelihoods from our $D_{A}(z)$ measurements which have been 
obtained from photometric redshift samples. We have used the prior information of the sound horizon 
scale and $w_{m}$ from Planck like other BAO studies. We believe that the photo-z error is subdominant
compared to the error that we get from cosmic variance. The DESI catalogue and the DECaLS catalogue 
share the same footprint and so the cosmic variance will be minimal, however, due to the photo-z 
uncertainty, the error on the $H_{0}$ value we obtain increases. It can be seen that 
our mean value of $H_{0}=67.59$ km/s/Mpc is well within $1\sigma$ of the 
Planck18 + BAO value.

\section{Discussion and conclusions}\label{sec:conclusion}
We provide a statistical methodology to extract cosmic distance information using BAO peaks 
only from the DECaLS DR8 LRG photometric galaxy sample. Common practice to extract the BAO 
peak from photometric redshift catalogues is by measuring the incomplete angular correlation 
function. In this manuscript, we make use of the wedge correlation function, wherein we split 
the sample 
into small wedges and include the BAO information from all the wedges in which they are still 
present ($\bar \mu<0.5$, above which there is noticeable contamination). 
Transverse cosmic distance $D_{A}(z)$ is measured with good precision for both the redshift 
samples giving us values of 
$D_{A}(\bar{z}=0.69) = 1499 \pm 77 \, \mathrm{Mpc} (r_{d}/r_{d,fid})$ and 
$D_{A}(\bar{z}=0.87) = 1680 \pm 109 \, \mathrm{Mpc} (r_{d}/r_{d,fid})$ with a fractional 
error of 5.14\% and 6.48\% respectively. The values that we obtain have been compared with the 
theoretical prediction for $D_{A}(z)$ as a function of redshift obtained using the cosmological 
parameters measured from \cite{Planck_2018} and are well within the $1\sigma$ region. 

We have also 
compared our results with the results of $D_{A}(z)$ obtained from other similar surveys (both 
spectroscopic and photometric) and find them to be consistent with each other. 
Since most radial information is contained at the
$\bar \mu>0.5$ bins which are contaminated by the photometric redshift uncertainty, 
we are not able to extract information on the radial cosmic distance. 
This is the first time that $D_{A}(z)$ is constrained at such a high redshift ($\bar{z}=0.87$)
using LRGs. 

Most of the recent works \citep{Sanchez_empirical,Carnero_2012,Seo_2012} 
have used the angular correlation function $w(\theta)$ to get 
cosmic distance measures using several narrow redshift slices.
Since radial binning blends data beyond what is induced by the photometric redshift error, 
the full information that is present is not utilised.  
Another important aspect that is often ignored when calculating $w(\theta)$ is the cross 
correlation between the different redshift bins used along with the  
complications that it brings with calculating the covariance matrix, i.e. 
the computing time increases with the number of bins in $\theta$ and number of redshift 
slices used. 

The full spectroscopy DESI galaxy catalogue will be available around 2025 and will cover 
the footprint observed by DECaLS, but with higher precision.  
Here we try to probe the cosmological signature imprinted in this photometric footprint map. 
Most BAO measurements \citep{Anderson_2012,Alam_2017,Chuang_2017} 
at low redshifts have supported the $H_0$ measurement by the Planck experiment, 
and it becomes interesting to see whether DESI will provide a similar 
result or not. By using the information obtained on the angular diameter distance from the DECaLS 
samples at the two median redshifts along with prior information of the sound horizon from Planck, 
we try to provide a precursor for the $H_{0}$ value expected from DESI. 
Although precise information of $H_{0}$ is not possible from photometric redshift catalogues, we
obtain a value of $H_{0}=67.59\pm5.52$ km/s/Mpc with a fractional error of 8.16\%.  
Our value of $H_{0}$ supports the $H_{0}$ measured by all other BAO results and is consistent 
with the $\Lambda$CDM model. 

\acknowledgments
We would like to thank Alfonso Veropalumbo for providing us specific details on the empirical fitting 
procedure using the MCMC analysis. We would also like to thank Behzad Ansarinejad for general 
discussions on the BAO. 
Data analysis was performed using the high performance computing cluster \textit{POLARIS} 
at the Korea Astronomy and Space Science Institute. 
This research made use of TOPCAT and STIL: Starlink Table/VOTable Processing Software 
developed by \citet{topcat} 
and also the Code for Anisotropies in the Microwave Background (CAMB) \citep{CAMB_1,CAMB_2}.
Srivatsan Sridhar would also like to thank Sridhar Krishnan, Revathy Sridhar and 
Madhumitha Srivatsan for their support and encouragement during this work.

The Photometric Redshifts for the Legacy Surveys (PRLS) catalog used in this paper was produced thanks to funding from the 
U.S. Department of Energy Office of Science, Office of High Energy Physics via grant DE-SC0007914.

The Legacy Surveys consist of three individual and complementary projects: the Dark Energy Camera 
Legacy Survey (DECaLS; NOAO Proposal ID \# 2014B-0404; PIs: David Schlegel and Arjun Dey), 
the Beijing-Arizona Sky Survey (BASS; NOAO Proposal ID \# 2015A-0801; PIs: Zhou Xu and Xiaohui Fan), 
and the Mayall z-band Legacy Survey (MzLS; NOAO Proposal ID \# 2016A-0453; PI: Arjun Dey). 
DECaLS, BASS and MzLS together include data obtained, respectively, at the Blanco telescope, 
Cerro Tololo Inter-American Observatory, National Optical Astronomy Observatory (NOAO); 
the Bok telescope, Steward Observatory, University of Arizona; and the Mayall telescope, 
Kitt Peak National Observatory, NOAO. The Legacy Surveys project is honored to be permitted 
to conduct astronomical research on Iolkam Du'ag (Kitt Peak), a mountain with particular 
significance to the Tohono O'odham Nation.

NOAO is operated by the Association of Universities for Research in Astronomy (AURA) under a 
cooperative agreement with the National Science Foundation.

This project used data obtained with the Dark Energy Camera (DECam), which was constructed by the
Dark Energy Survey (DES) collaboration. Funding for the DES Projects has been provided by the 
U.S. Department of Energy, the U.S. National Science Foundation, the Ministry of Science and 
Education of Spain, the Science and Technology Facilities Council of the United Kingdom, the 
Higher Education Funding Council for England, the National Center for Supercomputing Applications 
at the University of Illinois at Urbana-Champaign, the Kavli Institute of Cosmological Physics 
at the University of Chicago, Center for Cosmology and Astro-Particle Physics at the Ohio 
State University, the Mitchell Institute for Fundamental Physics and Astronomy at Texas A\&M 
University, Financiadora de Estudos e Projetos, Fundacao Carlos Chagas Filho de Amparo, 
Financiadora de Estudos e Projetos, Fundacao Carlos Chagas Filho de Amparo a Pesquisa do 
Estado do Rio de Janeiro, Conselho Nacional de Desenvolvimento Cientifico e Tecnologico and the 
Ministerio da Ciencia, Tecnologia e Inovacao, the Deutsche Forschungsgemeinschaft and the 
Collaborating Institutions in the Dark Energy Survey. The Collaborating Institutions are 
Argonne National Laboratory, the University of California at Santa Cruz, the University of 
Cambridge, Centro de Investigaciones Energeticas, Medioambientales y Tecnologicas-Madrid, 
the University of Chicago, University College London, the DES-Brazil Consortium, the 
University of Edinburgh, the Eidgenossische Technische Hochschule (ETH) Zurich, Fermi National 
Accelerator Laboratory, the University of Illinois at Urbana-Champaign, the Institut de 
Ciencies de l'Espai (IEEC/CSIC), the Institut de Fisica d'Altes Energies, Lawrence Berkeley 
National Laboratory, the Ludwig-Maximilians Universitat Munchen and the associated Excellence 
Cluster Universe, the University of Michigan, the National Optical Astronomy Observatory, the 
University of Nottingham, the Ohio State University, the University of Pennsylvania, the 
University of Portsmouth, SLAC National Accelerator Laboratory, Stanford University, the 
University of Sussex, and Texas A\&M University.

The Legacy Survey team makes use of data products from the Near-Earth Object Wide-field Infrared 
Survey Explorer (NEOWISE), which is a project of the Jet Propulsion Laboratory/California 
Institute of Technology. NEOWISE is funded by the National Aeronautics and Space Administration.

The Legacy Surveys imaging of the DESI footprint is supported by the Director, Office of Science, 
Office of High Energy Physics of the U.S. Department of Energy under Contract No. DE-AC02-05CH1123, 
by the National Energy Research Scientific Computing Center, a DOE Office of Science User Facility 
under the same contract; and by the U.S. National Science Foundation, Division of Astronomical 
Sciences under Contract No. AST-0950945 to NOAO.

\software{astropy \citep{astropy_2013,astropy_2018},
		  TOPCAT \citep{topcat}, emcee \citep{emcee}}

\appendix
\restartappendixnumbering

\section{Systematic contamination from stellar density}\label{sec:appendix1}
Systematic effects are often present in an imaging dataset such as the DECaLS dataset, which can 
lead to spurious fluctuations in the target density and in turn to changes in the shape of the 
redshift distribution.
One such major contribution towards systematic contamination in the data comes from correlations
with stellar density \citep{Decals_2019}. To check for this systematic effect, we compare our 
DECaLS LRG density and the density of the random catalogue with the density of Gaia stars. First, we 
convert the sky coordinates (RA and DEC) from our data and random catalogue into HEALPIX pixels 
using the same $nside=256$ as used for the Gaia stellar density maps. 
We then use the Pearson correlation coefficient (PCC) to assess the linear 
correlation between the two datasets. For two variables $X$ and $Y$, the PCC is defined as, 
\begin{equation}
    \rho_{X,Y} = \frac{cov(X,Y)}{\sqrt{cov(X,X)cov(Y,Y)}}
\end{equation}
where $cov(X,Y)$ is the covariance between $X$ and $Y$ across all pixels. We get a value of 
$\rho_{X,Y}=-0.0416$ for $X$ and $Y$ being the Gaia stellar density and DECaLS LRG density and 
$\rho_{X,Y}=-0.0443$ for $X$ and $Y$ being the Gaia stellar density and random catalogue density, 
which shows that there is almost no strong positive or negative correlation between the two 
datasets separately. 

\begin{figure}\label{fig:ezmock_vs_decals}
\centering
\subfloat{
\includegraphics[width=0.32\textwidth]{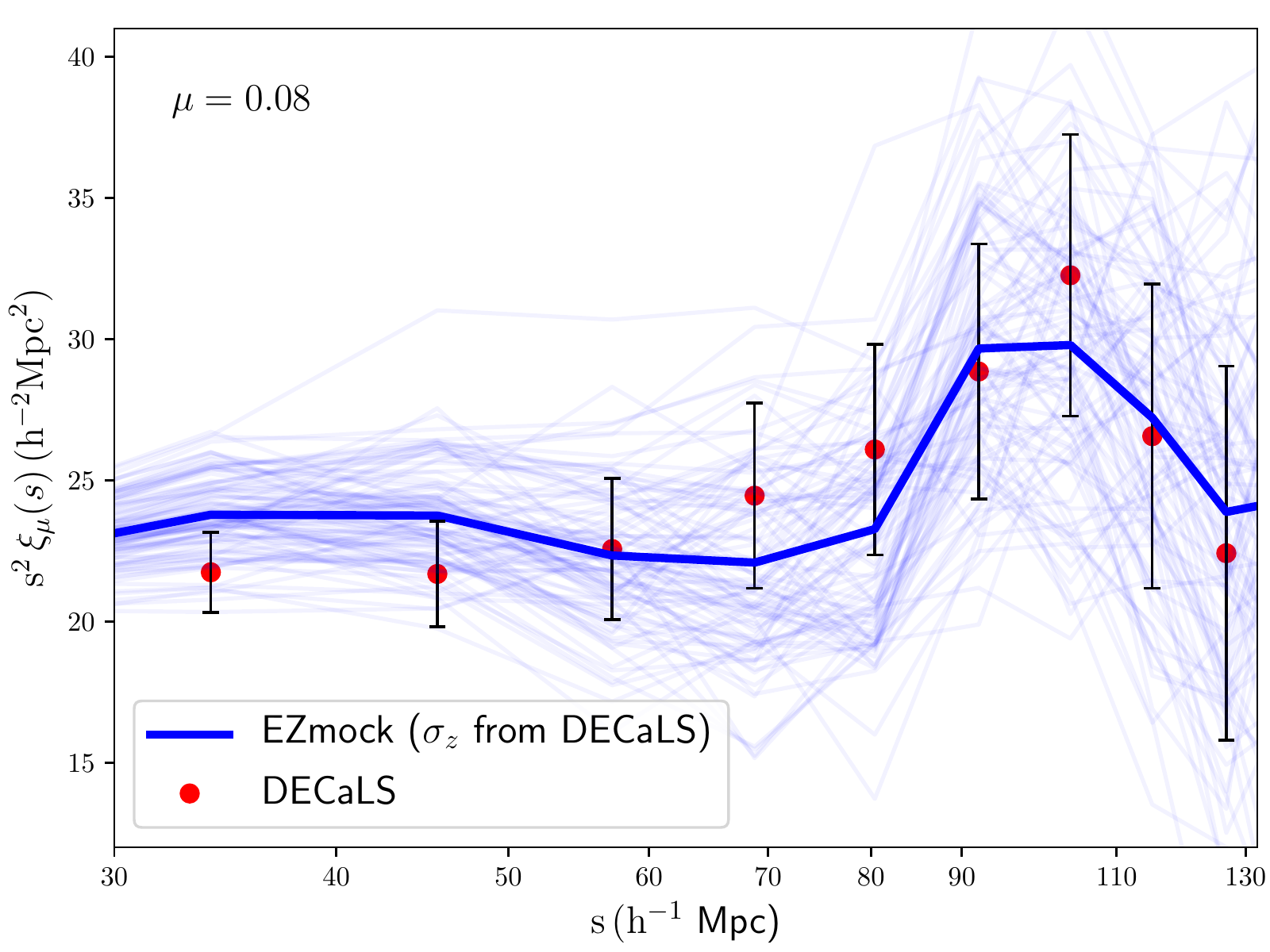}}\hfill
\subfloat{
\includegraphics[width=0.32\textwidth]{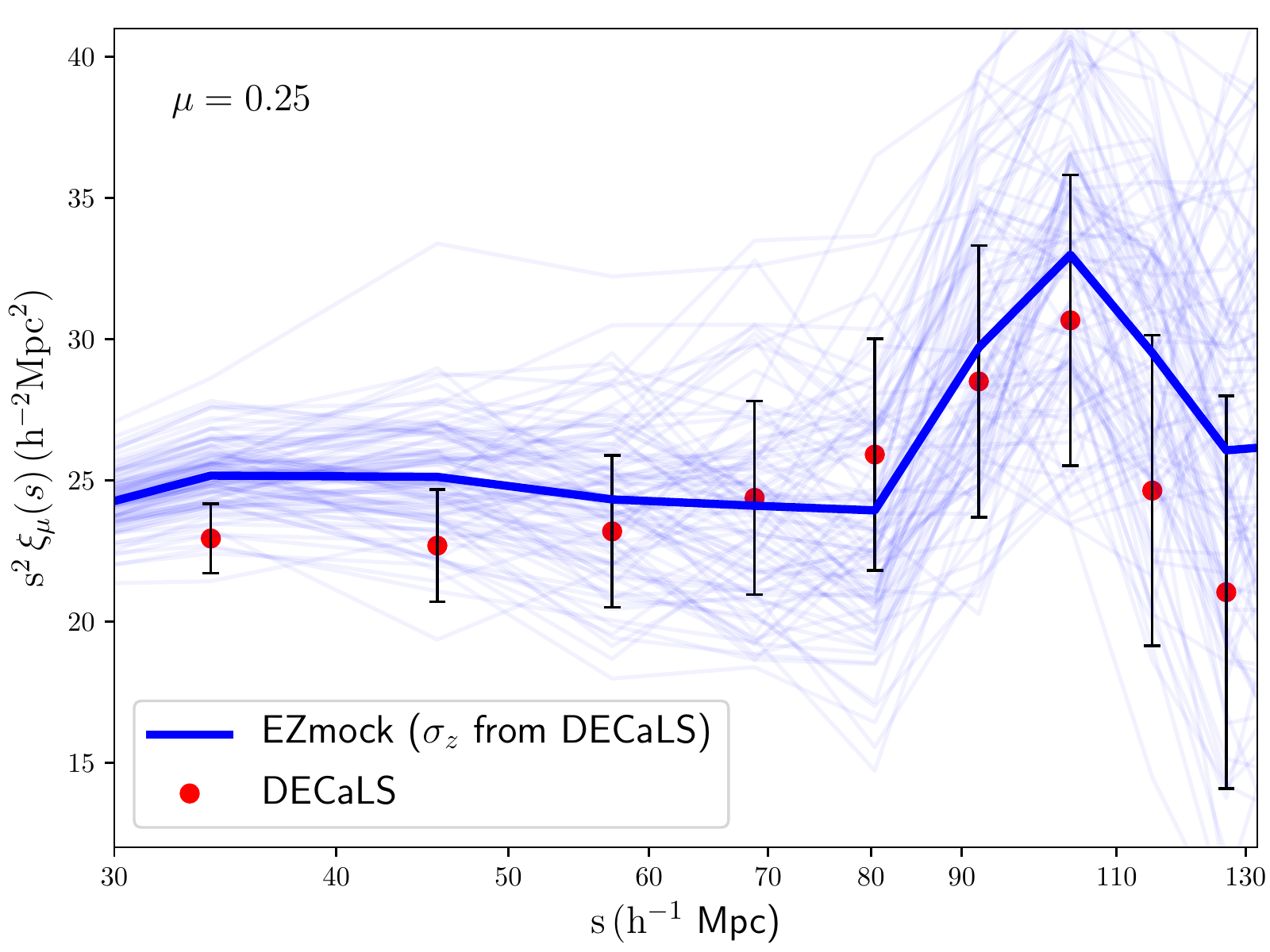}}\hfill
\subfloat{
\includegraphics[width=0.32\textwidth]{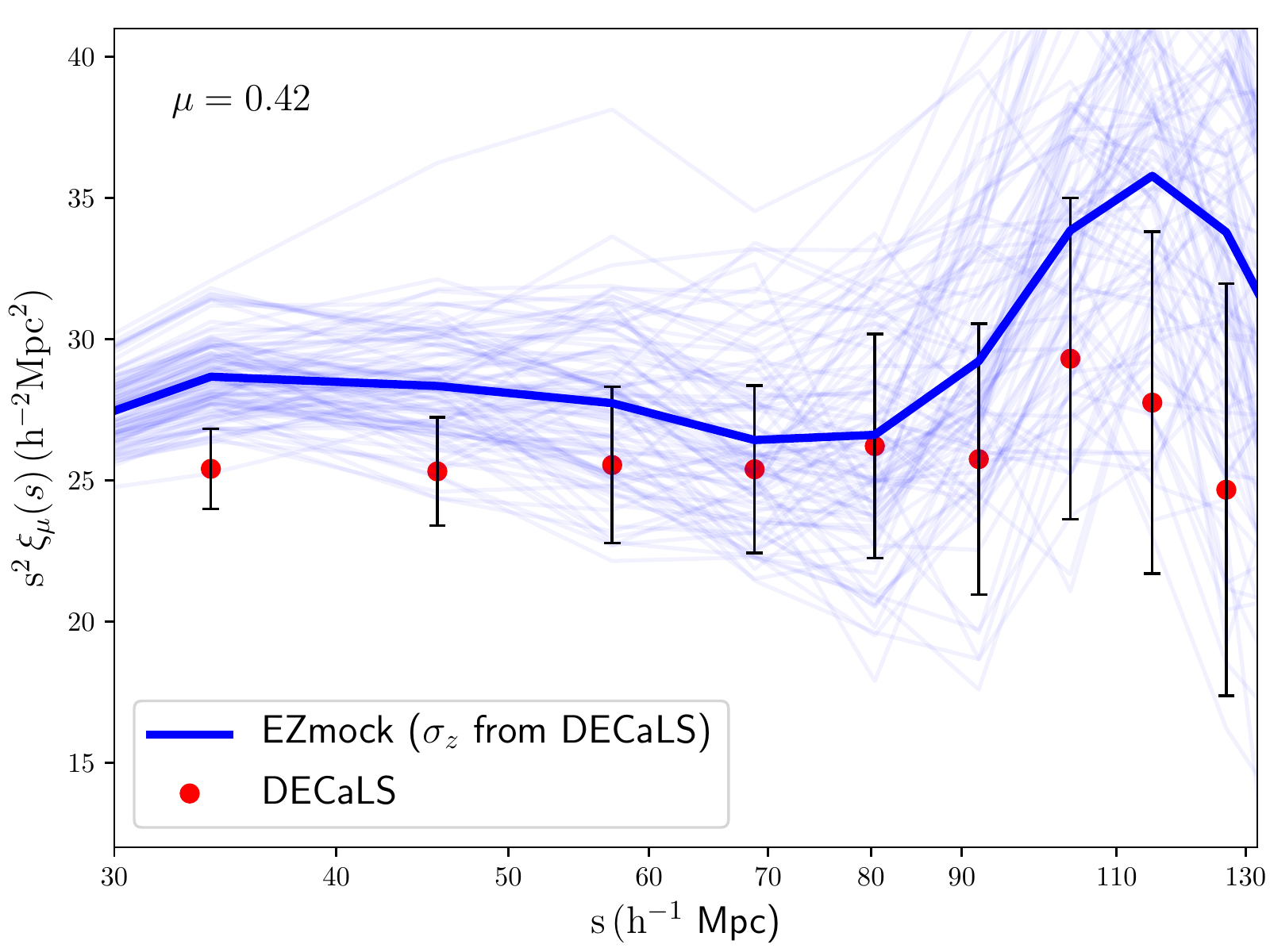}}\\
\subfloat{
\includegraphics[width=0.32\textwidth]{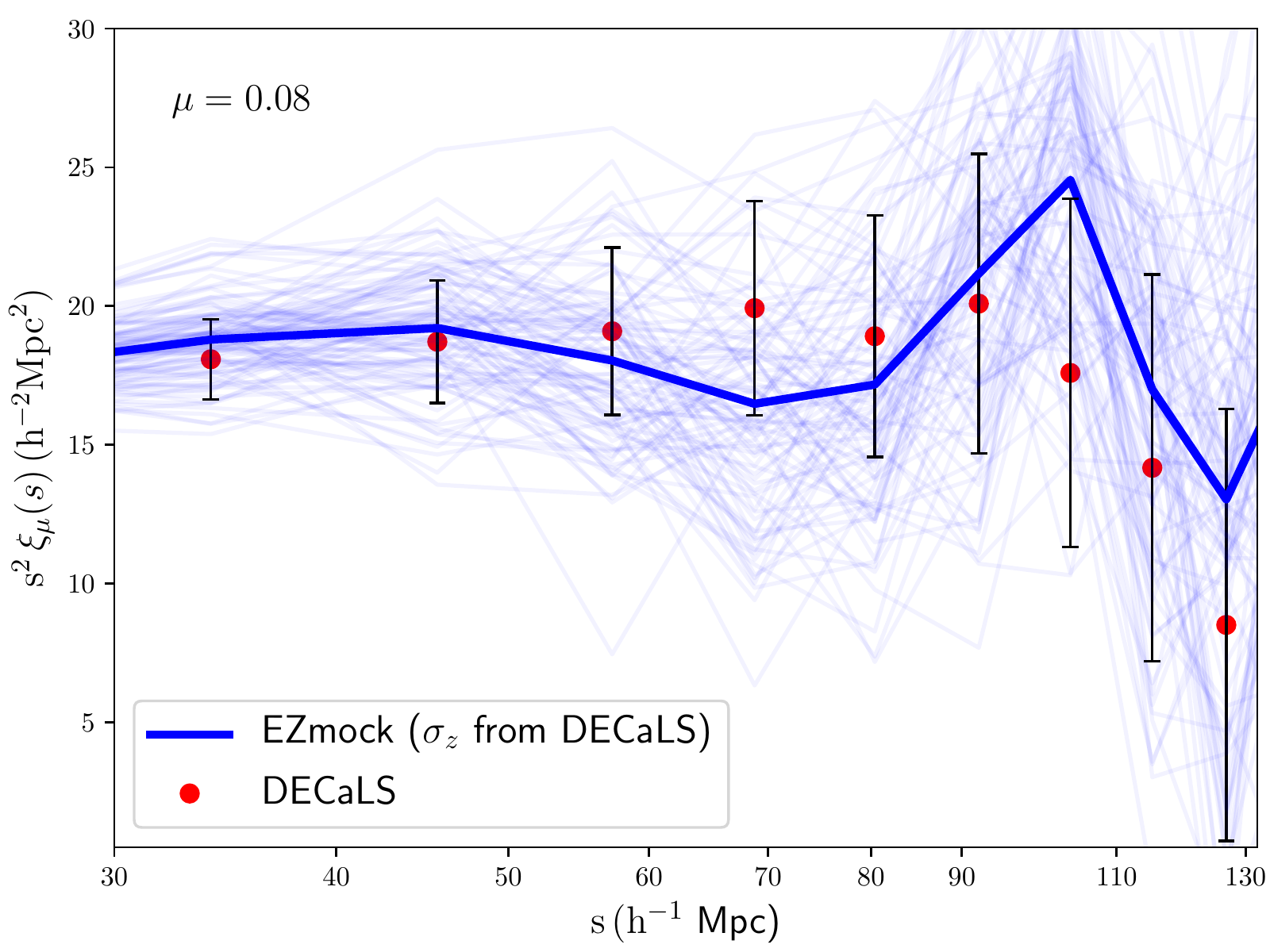}}\hfill
\subfloat{
\includegraphics[width=0.32\textwidth]{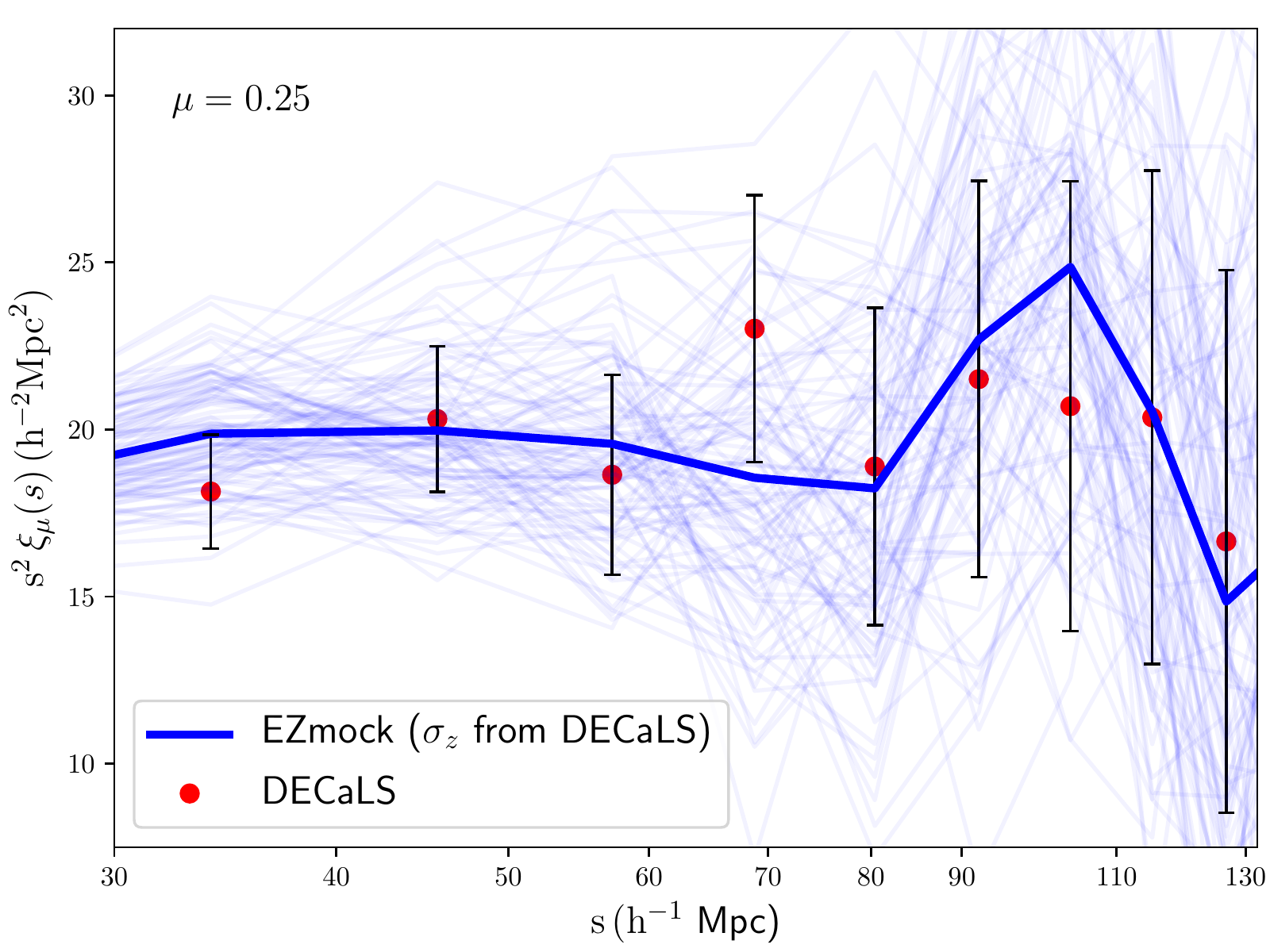}}\hfill 
\subfloat{
\includegraphics[width=0.32\textwidth]{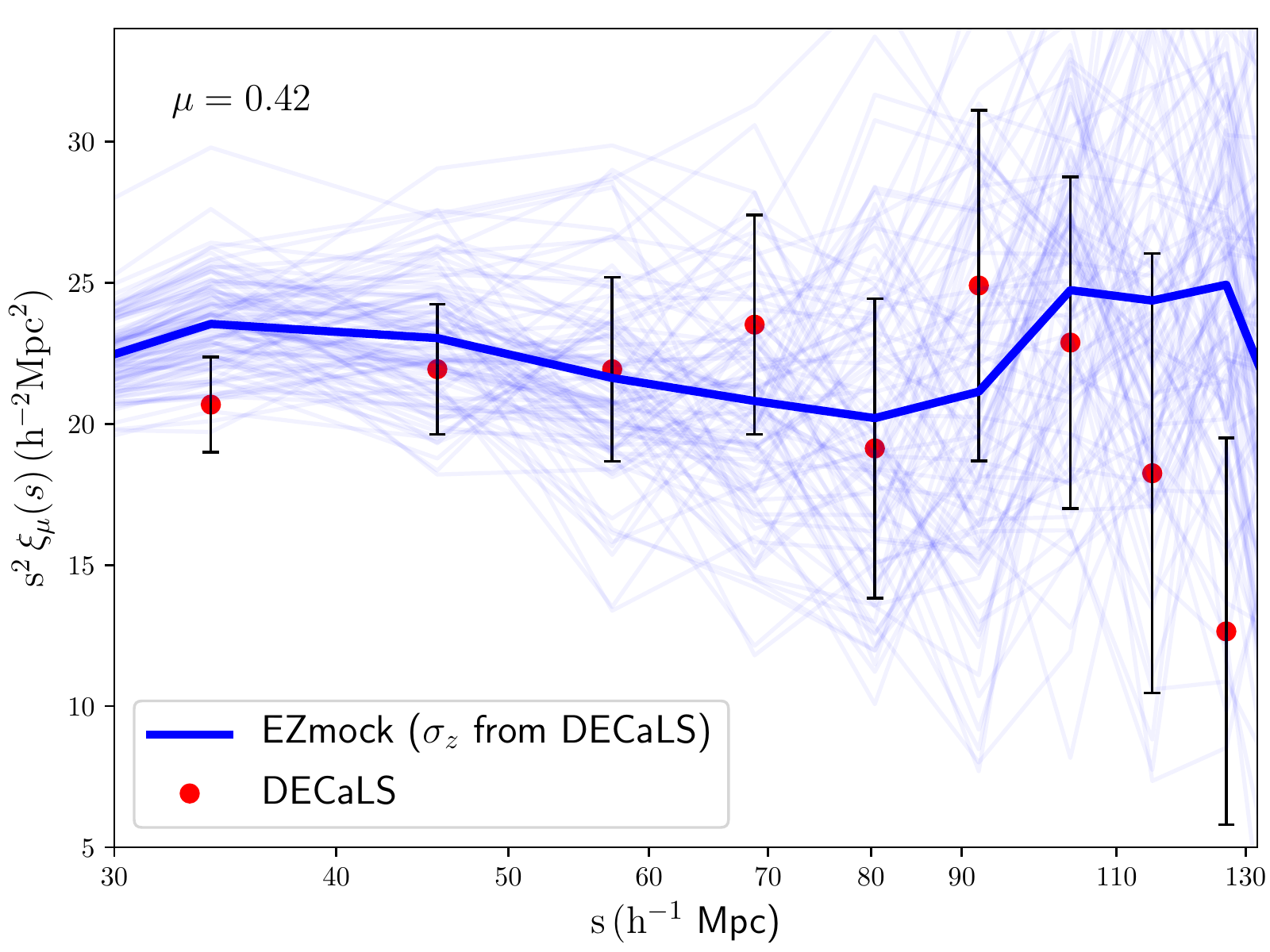}}
\caption{First row: Comparison of the correlation function \xismu (multiplied by $s^{2}$ ) 
         for $\bar \mu = 0.08,0.25,0.42$ (from the $0.6<z_{phot}<0.8$ sample) between the DECaLS data 
         (given by red dots) and the mean \xismu of the 100 EZmock samples (given by the solid blue line)
         generated by the Gaussian approximation method. 
         The photo-z's for all the EZmock samples have been created by extracting a random $\sigma_z$ 
         from the parent DECaLS sample. 
         The \xismu from all the 100 samples are plotted as lighter blue lines in the 
         background. Note: We have added a constant value of 0.0020 to the EZmock \xismu for the 
         $0.6<z_{phot}<0.8$ sample for the amplitudes to match. 
         Second row: The same as above, but for the $0.8<z_{phot}<1.0$ sample. We have added a 
         constant value of 0.0015 to the EZmock \xismu for the 
         $0.8<z_{phot}<1.0$ sample for the amplitudes to match. }
\label{fig:ezmock_diff_redshift_xismu}
\end{figure}

\section{Comparison of the clustering results between DECaLS and EZmock}\label{sec:appendix2}
As mentioned in Section \ref{sec:ezmock_data}, we generate photo-z's for the EZmock 
samples from random $\sigma_{z}$ values obtained from the DECaLS data by restricting to galaxies 
of similar redshifts. 
We calculate \xismu for the 100 EZmock samples separately using the same 
$s$ and $\mu$ binning scheme as mentioned in Section \ref{sec:decals_and_darksky_measurements} and 
the mean \xismu for both the redshift samples is plotted as the solid blue line in Figure 
\ref{fig:ezmock_diff_redshift_xismu}. We find that by using the true values of \xismu for the two 
redshift samples from the EZmock, the amplitudes of \xismu do not match. However, by adding adding a
constant value (0.0020 for the $0.6<z_{phot}<0.8$ sample and 0.0015 for the $0.8<z_{phot}<1.0$
sample) to the \xismu, we see that the mean \xismu of the obtained from the EZmock photo-z's
created by randomly selecting $\sigma_{z}$'s from the DECaLS sample match closely to the DECaLS \xismu,
especially at the BAO scales as shown in Figure \ref{fig:ezmock_diff_redshift_xismu}. To statistically 
measure the significance of the similarity, we use the KS test on both the redshift samples. 
The null hypotheses for the KS test is that the distributions are the same and to reject the 
null hypotheses we require a $p$-value (significance) less than 0.05. For 
the $0.6<z_{phot}<0.8$ sample, we obtain a $p$-value of 0.78 and 0.18 for the $\bar\mu=0.08$ and 0.25
samples respectively. For the $\bar\mu=0.42$ sample, we however obtain a $p$-value of 0.004 which rules
out the null hypotheses. We however, believe that this should not have a significant impact on the 
covariance matrix obtained and should be conservative. 
For the $0.8<z_{phot}<1.0$ sample, the minimum $p$-value we obtain from all the three $\bar\mu$ bins is
0.48, which does not rule out the null hypotheses. When adding the 
constant, we assume that there is some bias in the data, but we do not expect that it would change 
the covariance matrix. 

As a further test, we fit the mean \xismu from the 100 EZmock samples for the two redshift bins using
the MCMC technique with the same fitting parameter space as given by Eq.\ref{eqn:param_space}. 
We apply the same priors to all our samples as mentioned in Section \ref{sec:fitting_procedure}. 
The values of $s_{m}$ obtained from the MCMC fit for the two samples are given in Table 
\ref{tab:table3}. First we compare the $s_{m}$ value obtained from the EZmock samples with the same
obtained from the DECaLS data. It can be seen that the common trend of $s_{m}$ increasing with $\bar\mu$
bin is also observed for the EZmock sample in both the redshift ranges considered. The errors obtained
on $s_{m}$ for the EZmock samples are similar to what we have obtained for the DECaLS sample in both 
the redshift ranges, with the error on $s_{m}$ increasing with $\bar\mu$. 

The obtained $s_{m}$ values from the EZmock samples are converted to physical distances using the 
same methodology as explained in Section \ref{sec:dist_measures_final}. For the $0.6<z_{phot}<0.8$ 
and $0.8<z_{phot}<1.0$ samples we obtain values of 
$D_{A}(0.69)=1507\pm58\,\mathrm{Mpc}(r_{d}/r_{d,fid})$ and 
$D_{A}(0.87)=1616\pm84\,\mathrm{Mpc}(r_{d}/r_{d,fid})$. 
These values correspond to distance measures of 
3.88\% and 5.2\% precision for the two redshift samples respectively. It can also be noted that the recovered
values of $D_{A}$ for the EZmock samples are a good match to the expected values $D_{A}^{fid}$ from the fiducial 
cosmology we use. The fiducial values of $D_{A}^{fid}$ for our cosmology at the two mean redshifts is
$D_{A}^{fid}(\bar z=0.697) = 1514$ Mpc (0.4\% higher than the recovered value) and 
$D_{A}^{fid}(\bar z=0.874) = 1638$ Mpc (1.3\% higher than the recovered value). Given the overall precision, these 
values are well within the expected limits. 

\begin{table}
\begin{center}
\begin{tabular}{c c c } 
\hline\hline
Redshift range & $\bar{\mu}$ bin & $s_{m}$ (h$^{-1}$Mpc)  \\ [0.5ex] 
\hline
                   & 0.08 & $106.7^{+3.1}_{-3.5}$ \\ 
$0.6<z_{phot}<0.8$ & 0.25 & $111.4^{+3.9}_{-4.2}$ \\
                   & 0.42 & $114.4^{+5.9}_{-4.7}$ \\
\hline
                   & 0.08 & $107.6^{+5.0}_{-4.8}$ \\ 
$0.8<z_{phot}<1.0$ & 0.25 & $107.9^{+5.1}_{-4.3}$ \\
                   & 0.42 & $112.3^{+6.0}_{-5.1}$ \\
\hline
\hline
\end{tabular}
\caption{Results of fitting the EZmock correlation function for the two redshift samples and in the 3 
         $\bar{\mu}$ bins using Eq. \ref{eqn:empirical_fit}. The $s_{m}$ is the BAO peak point 
         obtained from the fit and the units are in h$^{-1}$Mpc.}
\label{tab:table3}
\end{center}
\end{table}

\bibliographystyle{aasjournal}
\bibliography{decals_dr7} 



\end{document}